\input harvmac
\input amssym.def
\input epsf.tex
\def\la{\lambda}    
\def\ga{\gamma}
\def\N{{\cal N}}   
\def\E{{\cal E}} 
\def\L{\Lambda}
\def\B{{\cal B}}    
\def\A{{\cal A}}
\def\H{{\cal H}}
\def\i{{\rm i}}
\def\g{{\frak g}}
\def\ie{{\it i.e.}}
\def\half{{1\over 2}}
\font\tenmsb=msbm10       \font\sevenmsb=msbm7
\font\fivemsb=msbm5       \newfam\msbfam
\textfont\msbfam=\tenmsb  \scriptfont\msbfam=\sevenmsb
\scriptscriptfont\msbfam=\fivemsb
\def\Bbb#1{{\fam\msbfam\relax#1}}

\def\Rop{{\Bbb R}}
\def\Zop{{\Bbb Z}}

\def\bbbc{{\mathchoice {\setbox0=\hbox{$\displaystyle\rm C$}\hbox{\hbox
to0pt{\kern0.4\wd0\vrule height0.9\ht0\hss}\box0}}
{\setbox0=\hbox{$\textstyle\rm C$}\hbox{\hbox
to0pt{\kern0.4\wd0\vrule height0.9\ht0\hss}\box0}}
{\setbox0=\hbox{$\scriptstyle\rm C$}\hbox{\hbox
to0pt{\kern0.4\wd0\vrule height0.9\ht0\hss}\box0}}
{\setbox0=\hbox{$\scriptscriptstyle\rm C$}\hbox{\hbox
to0pt{\kern0.4\wd0\vrule height0.9\ht0\hss}\box0}}}}

\def\figin{\epsfcheck\figin}\def\figins{\epsfcheck\figins}
\def\epsfcheck{\ifx\epsfbox\UnDeFiNeD
\message{(NO epsf.tex, FIGURES WILL BE IGNORED)}
\gdef\figin##1{\vskip2in}\gdef\figins##1{\hskip.5in}
\else\message{(FIGURES WILL BE INCLUDED)}%
\gdef\figin##1{##1}\gdef\figins##1{##1}\fi}
\def\DefWarn#1{}
\def\figinsert{\goodbreak\midinsert}
\def\ifig#1#2#3{\DefWarn#1\xdef#1{fig.~\the\figno}
\writedef{#1\leftbracket fig.\noexpand~\the\figno}%
\figinsert\figin{\centerline{#3}}\medskip\centerline{\vbox{\baselineskip12pt
\advance\hsize by -1truein\noindent\footnotefont{\bf Fig.~\the\figno:} #2}}
\bigskip\endinsert\global\advance\figno by1}

\lref\zhu{Y. Zhu, {\it Vertex operator algebras, elliptic functions
and modular forms}, J. Amer. Math. Soc. {\bf 9} (1996) 237.}

\lref\verlinde{E. Verlinde, {\it Fusion rules and modular
transformations in 2D conformal field theory}, Nucl. Phys. 
{\bf B300} (1988) 360.}

\lref\fss{J. Fuchs, B. Schellekens, C. Schweigert, {\it From Dynkin
diagram symmetries to fixed point structures},
Commun. Math. Phys. {\bf 180} (1996) 39; {\tt hep-th/9506135}.} 

\lref\kac{V.G. Kac, {\it Infinite dimensional Lie algebras}, Cambridge
University Press (1990) [3rd ed.].}

\lref\go{P. Goddard, D.I. Olive, {\it Kac-Moody and Virasoro algebras
in relation to quantum physics}, Int. Journ. Mod. Phys. {\bf A1}
(1986) 303.}

\lref\fuchs{J. Fuchs, {\it Affine Lie algebras and quantum groups},
Cambridge University Press (1992).}

\lref\gannon{T. Gannon, {\it Algorithms for affine Kac-Moody
algebras}, {\tt hep-th/0106123}.}

\lref\gannonf{T. Gannon, {\it Modular data: the algebraic
combinatorics of conformal field theory}, {\tt math.QA/0103044}.}

\lref\cardy{J.L. Cardy, {\it Boundary conditions, fusion rules and
the Verlinde formula}, Nucl. Phys. {\bf B324} (1989) 581.} 

\lref\PSS{G. Pradisi, A. Sagnotti, Y.S. Stanev, {\it Completeness   
conditions for boundary operators in 2d conformal field theory}, 
Phys. Lett. {\bf B381} (1996) 97; {\tt hep-th/9603097}.}

\lref\slansky{R. Slansky, {\it Group theory for unified model
building}, Phys. Rep. {\bf 79} (1981) 1.}

\lref\bppz{R.E. Behrend, P.A. Pearce, V.B. Petkova, J.-B. Zuber, 
{\it Boundary conditions in rational conformal field theories}, 
Nucl. Phys. {\bf B579} (2000) 707; {\tt hep-th/9908036}.}

\lref\fsc{J. Fuchs, B. Schellekens, C. Schweigert, {\it A matrix S
for all simple current extensions}, Nucl. Phys. {\bf B473} (1996) 323;
{\tt hep-th/9601078}.}

\lref\gw{T. Gannon, M.A. Walton, {\it On fusion algebras and modular
matrices}, Commun. Math. Phys. {\bf 206} (1999) 1; 
{\tt q-alg/9709039}.}  

\lref\oc{A. Ocneanu, {\it talk}, Kyoto (2000).}

\lref\pz{V.B. Petkova, J.-B. Zuber, {\it Boundary conditions in charge
conjugate sl(N) WZW theories}, {\tt hep-th/0201239}.}

\lref\gannonb{T. Gannon, {\it Boundary conformal field theory and
fusion ring representations}, {\tt hep-th/0106105}.}

\lref\dfz{P. Di Francesco, J.-B. Zuber, {\it SU(N) lattice
integrable models associated with graphs}, Nucl. Phys. {\bf B338}
(1990) 602.}

\lref\petzu{V. Petkova, J.-B. Zuber, {\it From CFT to graphs},
Nucl. Phys. {\bf B463} (1996) 161; {\tt hep-th/9510175}.}

\lref\gaberdiel{M.R. Gaberdiel, {\it Fusion of twisted
representations}, Int. Journ. Mod. Phys. {\bf A12} (1997) 5183; 
{\tt hep-th/9607036}.}

\lref\bev{J. B\"ockenhauer, D.E. Evans,  
{\it Modular invariants, graphs and $\alpha$-induction for nets of
subfactors. III}, Commun. Math. Phys. {\bf 205} (1999) 183; 
{\tt hep-th/9812110}.}  

\lref\fsgen{J. Fuchs, C. Schweigert, {\it Symmetry breaking
boundaries I. General theory},  Nucl. Phys. {\bf B558} (1999) 419; 
{\tt hep-th/9902132}.}

\lref\pezu{V.B. Petkova, J.-B. Zuber, {\it Conformal field theories,
graphs and quantum algebras}; {\tt hep-th/0108236}.}

\lref\bek{J. B\"ockenhauer, D.E. Evans, Y. Kawahigashi, {\it Chiral
structure of modular invariants for subfactors},
Commun. Math. Phys. {\bf 210} (2000) 733; {\tt math.oa/9907149}.}

\lref\xu{F. Xu, {\it New braided endomorphisms from conformal
inclusions}, Commun. Math. Phys. {\bf 192} (1998) 349.}

\lref\sy{A.N. Schellekens, S. Yankielowicz, {\it Extended chiral
algebras and modular invariant partition functions}, Nucl. Phys. 
{\bf B327} (1989) 673.}

\lref\mp{W.G. McKay, J. Patera, {\it Tables of dimensions, indices,
and branching rules for representations of simple Lie algebras},
Marcel Dekker Inc., New York (1981).}

\lref\afflud{I. Affleck, A.W. Ludwig, {\it Universal noninteger `ground
state degeneracy' in critical quantum systems}, Phys. Rev. Lett. 
{\bf 67} (1991) 161.}

\lref\intril{K. Intriligator, {\it Bonus Symmetry in Conformal
Field Theory}, Nucl. Phys. {\bf B332} (1990) 541.}

\lref\ishikawa{H. Ishikawa, {\it Boundary states in coset conformal
field theories}, {\tt hep-th/0111230}.}

\lref\fhsw{J. Fuchs, L. Huiszoon, B. Schellekens, C. Schweigert,
J. Walcher, {\it Boundaries, crosscaps and simple currents},
Phys. Lett. {\bf B495} (2000) 427; {\tt hep-th/0007174}.} 

\lref\solit{J. Fuchs, C. Schweigert, {\it Solitonic sectors,
alpha-induction and symmetry breaking boundaries}, Phys. Lett. 
{\bf B490} (2000) 163; {\tt hep-th/0006181}.} 

\lref\bfs{L. Birke, J. Fuchs, C. Schweigert, {\it Symmetry breaking  
boundary conditions and WZW  orbifolds}, Adv. Theor. Math. Phys. 
{\bf 3} (1999) 671; {\tt hep-th/9905038}.}

\lref\mue{M. M\"uger, {\it Conformal Field Theory and
Doplicher-Roberts Reconstruction}, Fields Inst. Commun. {\bf 30}
(2001) 297; {\tt math-ph/0008027}.}


\Title{\vbox{\baselineskip12pt
\hbox{hep-th/0202067}
\hbox{KCL-MTH-02-03}}}
{\vbox{\centerline{Boundary states for WZW models}}}
\smallskip
\centerline{Matthias R. Gaberdiel%
\footnote{$^\ast$}{{\tt mrg@mth.kcl.ac.uk}}} 
\smallskip
\centerline{\it Department of Mathematics, King's College London} 
\centerline{\it Strand, London WC2R 2LS, U.K.}
\bigskip
\centerline{and}
\bigskip
\centerline{Terry Gannon%
\footnote{$^\star$}{{\tt tgannon@math.ualberta.ca}}}
\smallskip
\centerline{\it Department of Mathematical Sciences, University of
Alberta} 
\centerline{\it Edmonton, Alberta, Canada, T6G 2G1}\bigskip
\medskip
\vskip1.5cm
\centerline{\bf Abstract}
\bigskip
\noindent
The boundary states for a certain class of WZW models are
determined. The models include all modular invariants that are
associated to a symmetry of the unextended Dynkin diagram. Explicit
formulae for the boundary state coefficients are given in each case,
and a number of properties of the corresponding NIM-reps are derived.    
\Date{February, 2002}

\newsec{Introduction}

Given a conformal field theory defined on closed Riemann surfaces (a
closed string theory) we can ask whether the theory can be defined on
Riemann surfaces with boundaries. More precisely we can ask which
boundary conditions can be imposed at the various boundaries. From a
string theory point of view, this is the question of  which open
strings can be consistently added to a given closed string theory.   
In general, rather little is known about when a `complete' set of
boundary conditions can be found. It is believed that for closed
string theories that are consistently defined on arbitrary genus
surfaces such a solution exists, but not even this is clear. Even if
the higher genus property is sufficient, it seems unlikely to be
necessary.   

One of the conditions that every consistent set of boundary states 
has to satisfy is the so-called Cardy condition \refs{\cardy}: suppose
that the conformal field theory is a rational conformal field theory
with respect to some symmetry algebra $\A$, which possesses $N$
irreducible highest weight representations. The Cardy condition then
implies that every set of $M$ boundary conditions determines $N$
$M\times M$ matrices of non-negative integers, one for each highest
weight representation of $\A$. Furthermore, if the set of boundary
conditions is complete in a suitable sense \refs{\PSS}, these  
matrices actually form a (Non-negative Integer Matrix) representation, 
or NIM-rep for short, of the fusion algebra (see
\refs{\bppz,\gannonf,\gannonb} for an introduction to these matters).

The condition of a representation of the fusion algebra to be a
NIM-rep is quite restrictive, and one can therefore attempt to
classify all such representations (irrespective of whether they arise
from boundary states or not). If modular invariance is indeed a
sufficient condition for the construction of the boundary states, the 
classification of NIM-reps then gives a restriction on the possible
modular invariant theories. This program has been performed for the
case of the WZW models corresponding to su(2) in \refs{\bppz}. However,
in general this approach does not seem to be powerful since there are
typically many more NIM-reps than modular invariant conformal field 
theories \refs{\gannonb}, the best known examples being the tadpoles
of su(2) \refs{\bppz}. (There are also many modular invariant 
partition functions that do not have a NIM-rep \refs{\gannonb};
however, at least some  of them do not define consistent conformal
field theories, and it is therefore conceivable that modular
invariance is indeed sufficient for the construction of a boundary
theory.) In the following we shall call a NIM-rep `physical' if
it is compatible with a modular invariant (see the end of section~2 
below). If the NIM-rep possesses in addition compatible structures
corresponding to the defect lines and  3j- and 6j-symbols (see for 
example \refs{\pezu}), we shall call it `fully realised'. Fully
realised NIM-reps can be obtained from a subfactor \refs{\bev,\oc}. 
\medskip

It is generally believed that the conformal field theory whose space
of states is described by the diagonal modular invariant 
defines a consistent conformal field theory that is defined on
arbitrary Riemann surfaces.\footnote{$^\star$}{In the diagonal
modular invariant, left- and right-moving representations occur in
conjugate pairs. See section~2 for a more detailed definition.}
Given this theory, we can obtain other
consistent modular invariant conformal field theories by applying a
(global) symmetry transformation to the left-moving degrees of
freedom, say. For example, for the case of the WZW models
corresponding to the simple Lie algebra $\bar{\g}$, any symmetry of
the (unextended) Dynkin diagram can be employed. Symmetries of the 
unextended Dynkin diagram exist for $\bar{\g}={\rm su}(n)$ and $E_6$,
where they correspond to `charge conjugation'; for
$\bar{\g}={\rm so}(2n)$, where they correspond to the `chirality flip';
and for $\bar{\g}={\rm so}(8)$ where there is in addition `triality'.

The boundary states of the usual diagonal modular invariant theory are
known in detail and explicitly (they are described by Cardy's formula 
\refs{\cardy}), but the boundary states of the theories that can be
obtained using one of the above symmetries are less well
understood. In this paper we shall make a proposal for the 
boundary states of these theories. Our construction is based on the
observation that the problem of constructing boundary states for these
theories is essentially equivalent to the problem of finding the
twisted boundary states of the original theory. These twisted
boundary states satisfy two sets of Cardy-like conditions, one coming 
from the overlaps between the twisted boundary states among
themselves, and one from the overlaps between twisted boundary states 
and the usual Cardy states. These two Cardy conditions combine to give
a rather powerful constraint on the twisted boundary states, that will
allow us to make a very suggestive proposal. We are also able to show
that the NIM-rep entries it defines are always integers, and we have 
checked explicitly for numerous examples, that these integers are in
fact non-negative. 
\medskip

{}From a mathematical point of view, the WZW models are particularly
interesting since many of its quantities can be given a natural
algebraic interpretation in terms of the associated affine untwisted
Kac-Moody algebra ${\g}$. For instance, the highest weight 
representations of the symmetry algebra $\A$ are labelled by the
integrable highest weights $\la\in P_+$ of ${\g}$. The fusion
rules  $N_{\la\mu}^\nu$ can be interpreted as Weyl-folded tensor
product coefficients of the underlying finite-dimensional algebra
$\bar{\g}$. And the matrix $S$ which diagonalises the fusion rules
describes the $\tau\mapsto-1/\tau$ transformation of the characters 
$\chi_\la(\tau)$ of ${\g}$. 

\noindent It is therefore natural to ask three questions about the
NIM-reps in WZW models. Namely, find algebraic interpretations for
\vskip4pt

{\rightskip=12 true mm \leftskip=17 true mm \noindent 
\hskip-16pt (a) the boundary state labels $x\in\B$, that is to say the
rows and columns of the NIM-reps $\N_\la$ (in the type III subfactor 
language \refs{\bev} these are the  $M-N$ sectors); \vskip4pt

\noindent \hskip-16pt (b) the NIM-rep coefficients themselves; and
\vskip4pt 

\noindent \hskip-16pt (c) the matrix $\psi$ diagonalising the matrices 
$\N_\la$, and expressing the boundary states as linear combinations of
the Ishibashi states. \vskip4pt

}

\noindent In this paper we suggest answers for the NIM-reps that arise
as described above, and that are therefore associated to the
symmetries $\omega$ of the (unextended) Dynkin  diagram of
$\bar{\g}$. In particular, \vskip4pt 

{\rightskip=12 true mm \leftskip=17 true mm \noindent \hskip-16pt
(a) the $x\in\B$ are integrable highest weights for the twisted affine
algebra ${\g}^\omega$, or equivalently the  highest weights of
$\omega$-twisted representations of ${\g}$; \vskip4pt

\noindent \hskip-16pt (b) the $\N_{\la x}^y$ are the coefficients of
the fusion of the $\omega$-twisted ${\g}$-representation $x$ with the
untwisted  ${\g}$-representation $\la$; and \vskip4pt

\noindent \hskip-16pt (c) $\psi$ is the matrix $S$ describing how the
characters of the twisted algebra ${\g}^\omega$ transform under 
$\tau\mapsto -1/\tau$.\vskip4pt 

}

\noindent The exponents $\mu\in\E$ can also be interpreted as the
highest weights of another twisted algebra, the so-called orbit
algebra $\check{\g}$. Just as the symmetries of the (extended) Dynkin
diagram of $\g$ give rise to simple currents and thus to symmetries and  
gradings of the fusion coefficients, we will find that the symmetries
of the Dynkin diagrams of $\g^\omega$ and $\check{\g}$ give rise to
symmetries and gradings of the NIM-rep coefficients.

We also interpret our answers to (a), (b) and (c) in terms
of data for certain affine untwisted algebras;  in particular we
can express our NIM-rep coefficients using ordinary fusion
coefficients, {\it proving} the integrality of our NIM-reps.  

Answers to these questions, which appear to be similar to ours 
(although they were obtained by independent arguments),
appear in work by Fuchs, Schweigert, and collaborators (see in
particular \refs{\bfs,\solit}). However, we feel that the fundamental
simplicity of the above algebraic descriptions, and the fact that one
can be completely explicit, is rather elusive in their work. In
particular, their description is in terms of an orbifold construction,
and this hides the simple underlying structure we exhibit in our
paper. They have also not yet shown that their conjectured NIM-reps
are in fact integral.   
\smallskip

Given this algebraic description of the NIM-reps, it seems plausible
that our construction may generalise further to symmetries that are
only symmetries of the extended Dynkin diagram (and that therefore
give rise to simple current modular invariants). This is illustrated in
section~6 where we construct the NIM-reps for the simple current modular
invariants of su$(p)$ with $p\geq 2$ prime 
(for $p=2$ we ignore the easy case  $k/2$ odd). As long as 
$p^2$ does not divide $k+p$ we can furthermore prove that the NIM-rep 
entries are indeed non-negative integers. A more complete description
of these `simple current' NIM-reps and their non-negativity and
integrality will be given elsewhere. 
These simple current NIM-reps have 
been addressed in \refs{\fhsw} in considerable generality, although
in our opinion 
their description is much less explicit than our section~6. They
also provide an argument for the  integrality and non-negativity of
the NIM-rep coefficients, but it depends on the validity of their
conjectured formula for the fusion coefficients of the chiral algebra
extended by simple currents, and this conjecture has not yet 
to our knowledge 
been rigorously established.  For recent work on a completely
different approach  to proving this integrality and non-negativity, see \refs{\mue}.
\medskip

Surprisingly little is known about explicit NIM-reps, even for the WZW
models.  The known NIM-reps include at least one corresponding to each
modular invariant of 
su(2) \refs{\dfz} and su(3) \refs{\dfz,\bev,\oc,\bfs,\bppz} --- see
\refs{\bppz} for the lists; at least one NIM-rep for each known
modular invariant
of su(4) \refs{\petzu,\bev,\oc,\bfs,\fhsw}; and a few NIM-reps for
su(5) and  $G_2$ \refs{\dfz,\petzu,\bev,\bfs,\fhsw}. The list of NIM-reps for
su(2) is complete \refs{\dfz}; the lists of `fully realised' NIM-reps
for su(3) and su(4) have also been claimed to be complete \refs{\oc},
although the arguments have not  yet appeared in print. On the other
hand, it is known that the lists of all NIM-reps (without assuming 
them to be `physical' or `fully realised') are incomplete for all
algebras other than su(2). As mentioned before, NIM-reps associated to
conjugation automorphisms and simple currents are constructed for instance in
\refs{\bfs,\fhsw}; however their formulae are far more complicated and
less explicit than what is done here and in other papers quoted above.
In the type III subfactor
framework, NIM-reps compatible with the modular invariants
corresponding to simple currents \refs{\bev} and conformal embeddings
\refs{\xu,\bev}, were shown to exist for su$(n)$, and thus these
NIM-reps are in principle computable from the subfactor machinery
\refs{\bek,\bev}, although this computation is not entirely
straightforward and has not been performed. The diagonalising matrix
$\psi$, however, is at present inaccessible from the subfactor
approach.  

While this paper was being completed, \refs{\pz} appeared in which 
the  NIM-reps corresponding to charge conjugation $C$ for su$(n)$ are 
obtained, using different methods than ours. (In particular, their
analysis does not involve twisted affine algebras.) Their 
NIM-reps seem to be consistent with ours, although their
diagonalising $\psi$-matrices differ from ours by some (immaterial)
signs, and their boundary state labels are different ({\it e.g.}
for su(5) level 2, their boundary states $(m_1,m_2)=(1,1),(1,3)$
and $(2,1)$ correspond to our labels $(a_1,a_2)=(0,1),(0,0)$ and
$(1,0)$, respectively). After completing  this paper we were also
made aware of the paper \refs{\ishikawa} whose results have some
overlap with our section~3.    
\medskip

The paper is organised as follows. In section~2 we describe our
conventions and some background material. Our construction of the
NIM-reps is motivated in section~3 where we explain how it can be
understood in terms of constructing twisted boundary conditions for
the original diagonal theory. In section~4 the NIM-reps are 
explicitly described, and their relation to ordinary fusion matrices 
is exhibited, thereby proving the integrality of the NIM-rep
coefficients. Details of the underlying calculations are given  
in section~5. In section~6 we describe the NIM-reps for some simple
current modular invariants, and prove that, under suitable conditions, their
entries are indeed non-negative integers. Finally, section~7 contains
some conclusions. We have included an appendix in which the graphs of
some of the NIM-reps are explicitly given for small rank and level.

\newsec{Conventions and background material}

\noindent We will be interested here in the rational conformal field 
theory associated to an affine untwisted Kac-Moody algebra
${\g}$ at positive integer level $k$. Let us denote by 
$P_+=P_+^k({\bar{\g}})$ 
the set of all (finitely many) integrable level
$k$ highest weight representations of ${\g}$. The corresponding
highest weights can be written as
$\la=\sum_{i=0}^n\la_i\Lambda_i=(\la_0;\la_1,\ldots,\la_n)$, where the
non-negative integers $\la_i$ are the Dynkin labels of $\la$, and the
$\Lambda_i$ are the fundamental weights of ${\g}$. For instance, the
vacuum representation corresponds to the weight $k\Lambda_0$.   

The characters $\chi_\la(\tau)$, $\la\in P_+$, of a rational conformal 
field theory carry a representation of the modular group \refs{\zhu},
and in particular 
\eqn\modular{
\chi_\la(-1/\tau)=\sum_{\mu\in P_+}S_{\mu\la}\, \chi_\mu(\tau)\,,}
where $S$ is a unitary symmetric matrix. By the Verlinde formula
\refs{\verlinde} the $S$-matrix determines the fusion rule
coefficients of the theory as  
\eqn\fusion{
N_{\la\mu}^\nu=\sum_{\kappa\in   P_+}
{S_{\la\kappa}S_{\mu\kappa}S_{\nu\kappa}^* \over S_{0\kappa}}\,.}
Charge conjugation $C$ is a permutation of $P_+$, sending the weight
$\la$ to the weight $\la^c$ contragredient to it; it is given by
$C=S^2$, and obeys $N_{\la\mu}^0=\delta_{\mu,C\la}$. $C$ always
corresponds to a symmetry of the unextended Dynkin diagram (although
this symmetry may be trivial).

Another permutation of $P_+$ which is of fundamental importance in the
theory, is due to the {\it simple currents} (see for example
\refs{\sy,\intril}). These are the weights $J\in P_+$ with quantum
dimension ${S_{J0}\over S_{00}}=1$; their fusion rules 
$N_{J, \mu}^\nu=\delta_{\nu,J\mu}$ define a permutation 
$\mu\mapsto J\mu$. This permutation obeys  
\eqn\simcur{
S_{\la\, J\mu}=e^{2\pi\i\,Q(\la)}\,S_{\la\mu}}
for a rational phase $Q(\la)$. For all WZW theories except
$\widehat{E}_8$ level 2, the simple currents correspond to symmetries 
of the extended Dynkin diagram. Simple currents give rise to
symmetries and gradings of fusion coefficients
\eqn\scfus{\eqalign{
N_{J\la,J'\mu}^{JJ'\nu}=&N_{\la\mu}^\nu\cr
N_{\la\mu}^\nu\ne 0\ \Longrightarrow&\ Q(\la)+Q(\mu)\equiv Q(\nu)\
({\rm mod}\ 1)\,.}}
We will encounter simple currents in our formulae in section~4 for the
NIM-rep coefficients, and more explicitly in section~6. 
\medskip

We shall only consider conformal field theories for which the left-
and right-moving chiral algebra is the same, namely the untwisted
Kac-Moody algebra ${\g}$. The space of states of the full
conformal field theory can then be decomposed into representations of
two copies of ${\g}$, and thus can be written as 
\eqn\full{
\H = \bigoplus_{\la,\mu\in P_+} M_{\la\mu} \,
\H_{\la} \otimes \overline{\H_{\mu}^\ast} \,,}
where $M_{\la\mu}$ are non-negative integers that describe the
multiplicity with which the various tensor products of representations
appear in $\H$. We shall call the theory defined by $M=I$ the
`diagonal' theory; from the point of view of constructing boundary
states, this is the simplest case (as we shall see momentarily). We
shall mainly be interested in theories that can be defined on the
torus; this requires, in particular, that the partition function
corresponding to \full,  
\eqn\partition{
Z(\tau)=\sum_{\la,\mu\in P_+} M_{\la\mu}\,
\chi_\la(\tau)\, \chi_\mu(\tau)^\ast \,,}   
is modular invariant. The easiest examples of modular invariants are
$M=I$, and $M=C$, where $C$ denotes charge conjugation. More
generally, any symmetry of the {\it unextended} Dynkin diagram defines
a modular invariant; on the other hand, symmetries of the 
{\it extended} Dynkin diagram (\ie\ simple currents) may or may not
yield a modular invariant for a given fixed level $k$ (for example $k$
must be even for su$(2)$). The symmetries of the unextended Dynkin
diagram are often called {\it conjugations}. 

For a given modular invariant $M$, we call the {\it exponents} of $M$ 
the diagonal terms $M_{\la\la}$, or more precisely, the `multi-set' 
$\E_M$ in which  $M_{\la\la}$ copies of primary field $\la$ appears,
for each $\la\in P_+$.
\bigskip

We shall be interested in constructing boundary states for these
conformal field theories. A boundary state $|\!| a \rangle\!\rangle$ 
is a coherent state in the full conformal field theory. In the
simplest case where the boundary preserves the full affine algebra
(that is generated by $J^b_m$) it is characterised by the `gluing
condition'   
\eqn\gluing{
\left( J^b_{m} + \bar{J}^b_{-m} \right) |\!| a \rangle\!\rangle =0
\qquad \hbox{for all $b$ and $m\in\Zop$}\,.}
Since the modes that appear in \gluing\ map each 
$\H_{\la} \otimes \overline{\H_{\mu}^\ast}$ into itself, we can solve
the gluing condition separately for each summand in \full. We can find
a non-trivial solution provided that $\H_{\la}$ is isomorphic to
$\H_{\mu}$, \ie\ for each $\la\in\E_M$. For each such $\la$, the
relevant coherent state $|\la\rangle\!\rangle$ is then unique up to
normalisation, and is called the {\it Ishibashi} state. We shall
always (partially) fix the normalisation of the Ishibashi states so
that  
\eqn\ishi{
\langle\!\langle \la |\, q^{\half(L_0+\bar{L}_0-{c\over12})}\,
|\mu\rangle\!\rangle = \delta_{\la\mu} \chi_\la(\tau) \,,
\qquad q=e^{2\pi i \tau}\,.}
Every boundary state can then be written as a linear combination of
Ishibashi states,  
\eqn\boundex{
|\!| a \rangle\!\rangle = \sum_{\mu\in\E_M} 
{\psi_{a\mu} \over \sqrt{S_{0\mu}}}
|\mu\rangle\!\rangle\,,}
where we have introduced the factor involving $S_{0\mu}$ for future
convenience. Given the above normalisation of the Ishibashi states,
the boundary states are determined in terms of the matrix
$\psi_{a\mu}$\footnote{$^\dagger$}{Actually, \ishi\ only determines  
the normalisation of the Ishibashi state up to an arbitrary
phase. Furthermore, if some $\mu$ appears with non-trivial
multiplicity $n$ in $\E_M$, we can redefine the Ishibashi states by a 
transformation in $U(n)$. Unless these choices are specified
further, there is therefore an ambiguity in the definition of
$\psi$. This accounts for the difference in our formulae for $\psi$ 
from those in \pz.}.    
In general we expect to be able to find as many boundary
states as there are Ishibashi states, \ie\ that $\psi$ is a square
matrix. Furthermore, the completeness argument of \refs{\PSS} suggests
that $\psi$ should be unitary.

Not every linear combination \boundex\ defines an actual boundary
state. The allowed boundary states have to satisfy a number of
consistency conditions, the most important of which is the so-called 
Cardy condition \refs{\cardy} which can be understood as follow. The
`overlap' between two boundary states can be calculated from \boundex\ 
as  
\eqn\overlap{
\langle\!\langle a |\!|\, q^{\half(L_0+\bar{L}_0-{c\over12})}\,
|\!|b\rangle\!\rangle = \sum_{\mu\in\E_M}
{\psi_{a\mu}^\ast \psi_{b\mu} \over S_{0\mu}}\, \chi_\mu(\tau)\,.}
Upon the modular transformation $\tau\mapsto -1/ \tau$, this amplitude
must then be expressible in terms of a non-negative integer
combination of characters, \ie\
\eqn\cardycon{
\sum_{\mu\in\E_M}
{\psi_{a\mu}^\ast \psi_{b\mu} \over S_{0\mu}}\, \chi_\mu(-1/\tau) = 
\sum_{\mu\in\E_M} \sum_{\lambda\in P_+} 
{\psi_{a\mu}^\ast \psi_{b\mu} \over S_{0\mu}} \,
S_{\lambda\mu} \, \chi_\lambda(\tau) =:
\sum_{\lambda\in P_+} \N_{\lambda b}^a\, \chi_\lambda(\tau) \,,}
where $\N_{\lambda b}^a$, defined by 
\eqn\Nimdef{
\N_{\lambda b}^a = \sum_{\mu\in\E_M} 
{\psi_{a\mu}^\ast\, S_{\lambda\mu}\, \psi_{b\mu} \over S_{0\mu}}\,,}
must be a matrix of non-negative integers. The collection of these
matrices is usually called a {\it NIM-rep}. More abstractly, we
define a NIM-rep $\N$ to be an assignment of a matrix $\N_\la$, with
non-negative integer entries, to each $\la\in P_+$ such that $\N$ forms
a representation of the fusion ring  
\eqn\nim{
\N_\la\,\N_\mu=\sum_{\nu\in P_+} N_{\la\mu}^\nu\,\N_\nu\,,}
for all primaries $\la,\mu,\nu\in P_+$. Furthermore we require that 
\eqn\nimone{\eqalign{
\N_0=&\,I \cr 
\N_{C\la}=&\,\N_\la^t\qquad \la\in P_+ \,.  }}
If the NIM-rep arises as in the above construction of boundary states
with $\psi$ being unitary, it is easy to see that \nim\ and \nimone\
follow from \Nimdef. The easiest example of a NIM-rep are the fusion
matrices themselves since the assignment $\la\mapsto N_\la$ clearly
satisfies both \nim\ and \nimone. 

The matrices $\N_\la$ of any NIM-rep can always be simultaneously
diagonalised, by a unitary matrix $\widehat{\psi}$, in such a way that  
\eqn\psiab{
(\N_\la)_{xy}=\N_{\la x}^y=\sum_{\mu\in\E}
{\widehat{\psi}_{x\mu}\,S_{\la\mu}\,
\widehat{\psi}^*_{y\mu}\over S_{0\mu}}\,.}
The sum will be over some `multi-subset' $\E=\E(\N)$ of $P_+$,
\ie\ each element of $P_+$ will come with a multiplicity, possibly
zero. This multi-set $\E$ is called the {\it exponents} of the NIM-rep
$\N$. The eigenvalues and simultaneous eigenvectors (\ie\ the columns
of $\widehat\psi$) of the matrices $\N_\la$ are parametrised naturally
by the exponents $\mu\in\E(\N)$. The entries of these eigenvectors,
\ie\ the rows of $\widehat\psi$, are parametrised by $x\in\B$; the
same labels also  run through the rows and columns of the matrices
$\N_\la$. 

As we shall see later on, some of the NIM-reps that we shall discuss
in this paper have an analogue of a simple current symmetry 
(compare \scfus). By this we mean permutations $J^B$ of $\B$ and
$J^E$ of $\E$, such that  
\eqn\scnim{\eqalign{
\widehat\psi_{a\, J^E\mu}=&e^{2\pi\i\,Q^E(a)}\,\widehat\psi_{a\mu}\cr
\widehat\psi_{J^B a\, \mu}=&
e^{2\pi\i\,Q^B(\mu)}\,\widehat\psi_{a\mu}\cr 
\N_{\la,J^Ba}^{J^Bb}=&\N_{\la a}^b\cr
\N_{\la a}^b\ne 0\ \Longrightarrow&\ Q^E(\la)+Q^E(a)\equiv Q^E(b)\
({\rm mod}\ 1)\,,}}
for appropriate phases $Q^B$ and $Q^E$.
\bigskip

Any modular invariant (boundary) conformal field theory has both a 
modular invariant $M$ and a NIM-rep $\N$. We say that the set of
boundary states is complete, if $M$ and $\N$ are compatible in the
sense that their exponents agree (including multiplicities):
$\E_M=\E(\N)$. If this is the case we can then also identify the
matrices $\psi=\widehat{\psi}$; this is to say, the diagonalising
matrix $\widehat{\psi}$ of the abstract NIM-rep has then an
interpretation in terms of the boundary states of the conformal field
theory.  

One example where this identification has been understood is the
diagonal modular invariant theory $M=I$ for which $\E_M=P_+$. The
boundary states of this theory have been constructed in \refs{\cardy};
they are labelled by elements in $P_+$ and defined by $\psi=S$. The
corresponding NIM-rep is then the fusion matrix, as follows from
\fusion, which again has $\E(\N)=P_+$.    

However, in general it is not known how to construct the NIM-rep that
corresponds to the other modular invariants. In this paper we want to
solve this problem, in the context of the affine algebras, for those
modular invariants that correspond to a symmetry of the unextended
Dynkin diagram. We will also get a natural interpretation 
for the `boundary state labels' $x\in\B$.

\newsec{The abstract construction}

\noindent As we mentioned in the previous subsection, the boundary
states for the diagonal modular invariant are known, and
they are given by 
\eqn\cardyboun{
|\!| \la \rangle\!\rangle = \sum_{\mu\in P_+} 
{S_{\la\mu} \over \sqrt{S_{0\mu}}} \,
|\mu\rangle\!\rangle \,.}
Suppose $\omega$ is any outer automorphism of ${\g}$ that arises from
an  automorphism of the unextended Dynkin diagram. We want to find the 
boundary states of the modular invariant based on $\omega$. The
exponents $\E_\omega$ of the corresponding NIM-rep are precisely the 
subset of $P_+$ consisting of those representations that are invariant
under the action of  $\omega$. The problem of finding the boundary
states for the modular invariant based on $\omega$ is therefore
essentially equivalent to finding the $\omega$-twisted boundary states
for the original diagonal modular invariant theory. A boundary state
is $\omega$-twisted if it satisfies the twisted gluing 
condition    
\eqn\gluingtwist{
\left( J^b_{m} + \omega\left(\bar{J}^b_{-m}\right) \right) 
|\!| a \rangle\!\rangle^\omega =0 \qquad \hbox{for all $b$ and
$m\in\Zop$}\,.} 
Twisted boundary states have been discussed before in
\refs{\fsgen}. We shall propose a simple description for these 
boundary states (and their corresponding NIM-rep) below.

Every twisted boundary state can be written as a linear superposition of
the $\omega$-twisted Ishibashi states $|\mu\rangle\!\rangle^\omega$,  
that can be defined for each $\mu\in\E_\omega$, \ie\ for each 
$\mu\in P_+$ that is invariant under the induced action of $\omega$,
$\omega^\ast(\mu)=\mu$. Again, the Ishibashi states are uniquely
determined by this condition up to normalisation, which we (partially) 
fix by demanding that 
\eqn\ishitwis{
{}^\omega\langle\!\langle \la |\, q^{\half(L_0+\bar{L}_0-{c\over12})}\,
|\mu\rangle\!\rangle^\omega = \delta_{\la\mu}\, \chi_\la(\tau) \,.}
Let us expand the $\omega$-twisted boundary states as 
\eqn\bounomega{
|\!| x \rangle\!\rangle^\omega = \sum_{\mu \in \E_\omega}
{\psi_{x\mu} \over \sqrt{S_{0\mu}}} |\mu\rangle\!\rangle^\omega\,,}
where $x$ is, at this stage, an abstract label for the different
twisted boundary states. The NIM-rep that is associated to 
the $\omega$-twisted boundary states is then simply given by 
\eqn\twisNIM{
\N_{\lambda x}^{y} = \sum_{\mu\in\E_\omega} 
{\psi_{y\mu}^\ast \, S_{\lambda\mu}\, \psi_{x\mu} \over S_{0\mu}} \,,}
as follows from the same calculation that led to \Nimdef. 

\noindent Now we come to describe the main idea of our
construction. The advantage of considering the problem in the context
of the diagonal modular invariant theory (rather than as the problem
of finding the boundary states corresponding to a different theory) is
that we do not only have the Cardy condition that requires that the
boundary states \bounomega\ among themselves lead to non-negative
integers as in \twisNIM. If the twisted boundary states \bounomega\
are consistent, they must also satisfy an appropriate Cardy condition
involving the overlap between one of the twisted boundary states, and
one of the original Cardy states \cardyboun. As we shall explain
momentarily, this constraint turns out to be very powerful, and will
allow us to make a very suggestive proposal for $\psi$. 

In order to analyse this constraint we have to determine the 
overlap between the boundary state $|\!| \lambda \rangle\!\rangle$ and the
boundary state $|\!| y \rangle\!\rangle^\omega$. Given our expansion
in terms of the relevant Ishibashi states this can be easily done, and
we find 
\eqn\overlapone{
{}^{\omega} \langle\!\langle y |\!|\, 
q^{\half(L_0+\bar{L}_0-{c\over12})}\, |\!|\lambda\rangle\!\rangle =
\sum_{\mu\in\E_\omega} 
{\psi_{y\mu}^\ast\, S_{\lambda\mu} \over S_{0\mu}}
\chi_\mu^{(\omega)}(\tau)
\,,}
where $\chi_\mu^{(\omega)}(\tau)$ is the {\it twining character} in
the representation $\mu$, \ie\
\eqn\twining{
\chi_\mu^{(\omega)}(\tau) = 
{\rm Tr}_{\H_\mu}\left(\tau_\omega q^{L_0-{c\over 24}}\right)\,.}
Here $\tau_\omega$ is the induced action of $\omega$ on $\H_\mu$.

We can now use the fact that the above twining character agrees
precisely with the ordinary character of the so-called orbit Lie
algebra $\check{\g}$ \refs{\fss}. (Explicit descriptions for these Lie
algebras are given in table (2.24) of \refs{\fss}; see also the
translation of their notation to the notation of Kac \refs{\kac} on
page 12.)  

In order to relate the `closed string result' \overlapone\ to the `open
string picture' we have to apply again a modular transformation, as
before in \cardycon. In the present context this involves the
characters of $\check{\g}$ that do not transform into one another under
the $S$-modular transformation. Instead, the characters of $\check{\g}$
transform into linear combinations of characters of the twisted Lie
algebra ${\g}^\omega$. Let us label the representations of
${\g}^\omega$ by $\hat{l}$. Then the relevant formula is  
\eqn\modtwis{
\chi_{\mu}^{(\omega)}(-1/\tau) = \sum_{\hat{l}} 
\hat{S}_{\hat{l} \mu}\, \chi_{\hat{l}}(\tau)\,,}
where $\hat{S}$ is the corresponding (unitary) $S$-matrix that is
given in Thms.13.8, 13.9 of \refs{\kac} (except that for reasons of 
convenience we have chosen ours to be the transpose of Kac's). 
Every (ordinary) representation of
the twisted algebra ${\g}^\omega$ is, by definition, the same as a 
$\omega$-twisted representation of the original algebra ${\g}$. Thus 
$\hat{l}$ labels equally the $\omega$-twisted representations of
${\g}$ (see for example \refs{\go} for an introduction into these
matters). 

\noindent Putting all of this together, we can therefore write the
overlap \overlapone\ in the `open string picture' as 
\eqn\overlaptwo{
\sum_{\mu\in\E_\omega} \sum_{\hat{l}}
{\psi_{y\mu}^\ast\, S_{\lambda\mu} \,\hat{S}_{\hat{l}\mu} \over S_{0\mu}}
\, \chi_{\hat{l}} (\tau)\,.}
The relevant `Cardy' condition is then that
\eqn\cardytwis{
\N_{\hat{l} \lambda}^{y} = \sum_{\mu\in\E_\omega} 
{\psi_{y\mu}^\ast \, S_{\lambda\mu}\, \hat{S}_{\hat{l}\mu} \over S_{0\mu}}} 
defines non-negative integers. 
\smallskip

Next we want to make a suggestive proposal for how to solve the two 
`Cardy' conditions  \twisNIM\ and \cardytwis. The NIM-rep we are
looking for has dimension $|\E_\omega|$; because $\hat{S}$ is an
invertible matrix, this equals the number of $\omega$-twisted
representations of the conformal field theory. Now we can always
construct {\it a} NIM-rep of this dimension by taking $\B$ to label
the different $\omega$-twisted (irreducible highest weight)
representations, and by taking $\N_{\lambda x}^{y}$ to be the twisted
fusion rules that describe the fusion of the untwisted representation
$\lambda$ with the twisted representation $x$, leading to twisted
representations described by $y$. (The fusion of twisted
representations is for example discussed in \refs{\gaberdiel}.) These
twisted fusion rules automatically define a NIM-rep since fusion is
associative.  

This suggests therefore that the label $x$ that describes the twisted
boundary states can be identified with the label for the
$\omega$-twisted  representations. Furthermore, the two sets of
integers \twisNIM\ and \cardytwis\ should both simply be the twisted 
fusion rules.\footnote{$^\ddagger$}{After this paper was completed we
learned that the possibility of this interpretation in terms of
twisted fusion rules was made in passing in \refs{\solit}.} 
Since these fusion rules are symmetric we therefore
have, writing $\hat{l}=x$, 
\eqn\argumentone{
\N_{x \lambda}^{y} = \N_{\lambda x}^{y} \,.}
Comparing \twisNIM\ and \cardytwis\ this then implies that 
\eqn\central{
\psi_{x\mu} = \hat{S}_{x\mu}\,.}
Thus we propose that {\it the boundary states of the NIM-rep
associated to $\omega$ are labelled by the $\omega$-twisted
representations, and the NIM-rep itself is precisely described by
the twisted fusion rules. The matrix diagonalising the matrices
$\N_\lambda$ is the modular matrix $\hat{S}$.} 

\smallskip

Returning to the original problem of finding the boundary states for
the conjugation modular invariant, we can now write down a complete
set of boundary states, all of which satisfy the gluing condition 
\gluing. As above, these boundary states are labelled by the
$\omega$-twisted representations of $\g$, and they are given as 
\eqn\bounfinal{
|\!| x \rangle\!\rangle = \sum_{\mu \in \E_\omega}
{\hat{S}_{x\mu} \over \sqrt{S_{0\mu}}} 
|\mu\rangle\!\rangle\,,}
where $|\mu\rangle\!\rangle$ are the standard Ishibashi states that
exist for $\mu\in\E_\omega$. The Cardy condition for these boundary
states is now precisely the condition that \twisNIM\ are non-negative 
integers. 

Finally, we should mention that the above arguments imply that we have
a {\it generalised Verlinde formula}  
\eqn\argumenttwo{
\N_{x \lambda}^{y} = \sum_{\mu\in\E_\omega}
{\hat{S}_{y\mu}^\ast\, S_{\lambda\mu} \, \hat{S}_{x\mu} \over S_{0\mu}}
\,,}  
that describes the twisted fusion rules in terms of the $S$-matrices
$S_{\la\mu}$ and $\hat{S}_{x\mu}$.

\subsec{Generalised fusion algebras}

It is natural to think of our construction in terms of some
generalised fusion algebra that combines untwisted and twisted
representations. Let us concentrate on the case where the twist is of
order two in the following. The fusion matrix corresponding to an
untwisted representation $\lambda$ then takes the form    
\eqn\untwistedfusion{
N_\lambda^{\rm full} = 
\pmatrix{ N_{\lambda\mu}^\nu & 0 \cr 0 & \N_{\lambda x}^{y}} \,,}
while the fusion matrix for a twisted representation $x$ is 
\eqn\twistedfusion{
N_x^{\rm full} = 
\pmatrix{ 0 & \N_{x \mu}^{y} \cr \N_{x y}^{\mu} & 0 } \,.}
In both cases we have written the matrices in blocks corresponding to 
untwisted ($\la,\mu,\nu\in P_+$) and twisted ($x,y\in\B$)
representations. These fusion rules define a consistent fusion algebra
in the sense of \refs{\gannonf} (although, unlike the more familiar
fusion rings of conformal field theory, it is not {\it self-dual}).   

The NIM-rep that is defined by combining the untwisted and the twisted 
boundary states is precisely this generalised fusion algebra. In
particular, this implies that the NIM-reps corresponding to $M=I$ and
the one corresponding to the modular invariant associated to 
the conjugation $\omega$ 
combine into some larger algebraic structure. This assumption played
an essential role in our `derivation' above.

This is quite analogous to the manner in which group theorists treat
projective representations of a finite group $G$: they are interpreted
as true representations of an extension $H$ of the Schur multiplier 
$M(G)$ of $G$ by $G$. It would be interesting to try to push this
analogy further and see, for example, what the Schur multiplier
corresponds to in our case.
\smallskip

Since the algebra defined by the matrices \untwistedfusion\ and 
\twistedfusion\ defines a fusion algebra in the sense of
\refs{\gannonf}, it possesses a (non-symmetric) $S$-matrix that
diagonalises the fusion rules, and recovers the full fusion
coefficients $N^{\rm full}$ via Verlinde's formula. In the present
context this $(|P_+|+|\E_\omega|)\times (|P_+|+|\E_\omega|)$ 
$S$-matrix is simply given by 
\eqn\Sfull{
S^{\rm full}=\pmatrix{
   {1\over\sqrt{2}}S_{i\mu}   & S_{ij} \cr
   \pm{1\over\sqrt{2}} \hat{S}_{x\mu }&0     }\,,}
where the first $2|E_\omega|$ columns are parametrised by 
$\mu\in\E_\omega$ and either choice of sign (so each entry
`${1\over\sqrt{2}}S_{i\mu}$' appears twice), and where the final
$|P_+|-|\E_\omega|$ columns are parametrised by the  
$j\in P_+\setminus\E_\omega$. The first $|P_+|$ rows are 
parametrised by  $i\in P_+$, and the last $|\E_\omega|$ rows are 
parametrised by the $\omega$-twisted representations $x$ of ${\g}$.  
If we introduce the column vectors
\eqn\columnv{S^{\rm full}_{\updownarrow,(\mu,\pm)}=
\left(\matrix{S_{\updownarrow,\mu}\cr
\pm \hat{S}_{\updownarrow,\mu}}\right)\,, \qquad
S^{\rm full}_{\updownarrow,j}=
\left(\matrix{S_{\updownarrow,j}\cr 0}\right)\,,}
then we can rephrase this as follows: for each $\mu$ and choice of
sign, the column vector $S^{\rm full}_{\updownarrow,(\mu,\pm)}$
is an eigenvector of $N^{\rm full}_\lambda$ ($\lambda\in P_+$) with
eigenvalue 
${S_{\lambda\mu}\over S_{0\mu}}= 
S^{\rm full}_{\lambda,(\mu,\pm)} / S^{\rm full}_{0,(\mu,\pm)}$,
and an eigenvector of $N_x^{\rm full}$ ($x\in\B$) with eigenvalue 
$\pm{\hat{S}_{x\mu}\over S_{0\mu}}=S^{\rm full}_{x,(\mu,\pm)}/
S^{\rm full}_{0,(\mu,\pm)}$. Similarly, the column vector
$S^{\rm full}_{\updownarrow,j}$ is an eigenvector of 
$N^{\rm full}_\lambda$ ($\lambda\in P_+$) with 
eigenvalue ${S_{\lambda j}\over S_{0j}}=
S^{\rm full}_{\lambda j}/ S^{\rm full}_{0j}$, and
an eigenvector of $N_x^{\rm full}$ ($x\in\B$) with eigenvalue
$0=S^{\rm full}_{xj}/ S^{\rm full}_{0j}$.

This generalised fusion algebra also appears in \refs{\bfs} under the 
name `classifying algebra'. Their equations (4.5) and (4.7)
correspond here to \Sfull\ and \untwistedfusion, \twistedfusion,
respectively.

One consequence of \Sfull\ and Perron-Frobenius theory is that, if the
coefficients $\N_{\la x}^y$ are all to be non-negative, the entries 
$\hat{S}_{x 0}$ must be of constant sign for all $x\in\B$. We will see
in the next section that they are indeed always positive. This can be
regarded as a non-trivial test of non-negativity --- the only property
of our NIM-reps which we cannot prove in general. The property of
$\hat{S}_{x0}$ to be positive has also a physical
interpretation since $\psi_{x0}=\hat{S}_{x0}$ is proportional to the
(positive) boundary entropy of the boundary condition described by $x$ 
\refs{\afflud}.

\subsec{Generalised NIM-reps}

Defect lines (that is, non-local operators attached to
non-contractible loops) can be inserted into the partition functions 
of the torus and cylinder, resulting in two other sets of NIM-rep-like
data associated to a conformal field theory \refs{\pezu}. These also
appear naturally in the subfactor context (see for example Thm.~4.16
in  \refs{\bek}). In particular, to each pair $\la,\la'$ of weights in 
$P_+$, we should have a matrix 
$\widetilde{V}_{\la\la'}=(\widetilde{V}_{\la\la';i}^j)_{i,j\in\tilde{\B}}$ 
of non-negative integers, giving a representation of the
`double-fusion algebra'
\eqn\dbln{\widetilde{V}_{\la\la'}\,\widetilde{V}_{\mu\mu'}=
\sum_{\nu,\nu'\in P_+}
N_{\la\mu}^\nu N_{\la'\mu'}^{\nu'}\,\widetilde{V}_{\nu\nu'}\ ,}
and obeying both 
$\widetilde{V}_{\la,C\la';\tilde{1}}^{\tilde{1}}=M_{\la\la'}$
and $\widetilde{V}_{\la\la'}^t=\widetilde{V}_{C\la,C\la'}$. Here $M$
is the torus partition function of the theory, $C$
denotes charge conjugation, and $\widetilde{V}^t$ is the transpose of
$\widetilde{V}$. The label $\tilde{1}$ is a special element of
$\tilde{\B}$. These conditions  imply that the matrices
$\widetilde{V}_{\la\la'}$ can be simultaneously diagonalised by some
matrix $\widetilde{\psi}$, and that the eigenvalues of
$\widetilde{V}_{\la\la'}$ will be 
${S_{\la\mu}\, S_{\la'\mu'}\over S_{0\mu}\, S_{0\mu'}}$; 
its multiplicity is required to be $(M_{\mu\mu'})^2$. The other NIM-rep-like
data concerns a representation by non-negative integer matrices
$\widetilde{\N}_i=(\widetilde{\N}_{ix}^y)_{x,y\in\B}$, 
$i\in\tilde{\B}$, of a (not necessarily commutative) fusion-like ring
built out of the matrix $\widetilde{\psi}$ --- see section 4.2 of
\refs{\pezu} for details. 

In the case where $M$ is a permutation matrix
$M_{\la\mu}=\delta_{\mu,\pi\la}$ (this is, in particular, the case
when $M$ is a conjugation modular invariant, as considered in this
section, or the simple current modular invariant considered in section
6.1 below), it is easy to find these matrices
$\widetilde{V}_{\la\la'}$ and $\widetilde{\N}_i$: in particular, we
can naturally identify $\tilde{\B}$ with $P_+$, and put 
$\widetilde{V}_{\la\la'}=N_\la N_{\pi\la'}$ and
$\widetilde{\N}_\la=\N_\la$. Thus these other NIM-rep-like data can
also be constructed for the theories considered in this paper.

\newsec{Explicit formulae}

\noindent In this section we shall collect explicit formulae
describing the NIM-reps for the various classes of algebras. The 
NIM-rep is uniquely determined once we know the exponents $\E(\N)$,
the eigenvalues ${S_{\la\mu} \over S_{0\mu}}$ for all weights 
$\la\in P_+$ and exponents $\mu\in\E(\N)$, and the diagonalising
matrix $\psi_{x\mu}$. As we have explained above, the matrix $\psi$
can be identified with the modular $S$-matrix of the twisted algebra
${\g}^\omega$. In principle these $S$-matrices are known \refs{\kac},
but as for the case of the more familiar untwisted $S$-matrices, the
usual formulae are rather complicated (involving sums over Weyl
groups). In the following we shall therefore give simple expressions
for both $S$-matrices, following the ideas of \refs{\gannon}. We shall
also explain how the NIM-rep coefficients can be expressed in terms
of the usual fusion coefficients; this will make manifest that they
are integers, as they must be. We conjecture
 that they are also non-negative, and have
verified this on the computer for several algebras and levels. The
arguments behind these calculations are given in the following
section, and some fusion graphs are exhibited in the appendix. In the 
following we shall analyse the different algebras case by case; we
shall always use the same numbering of nodes as in
\refs{\kac,\slansky}.

\subsec{The A-series}
 
\noindent A level $k$ weight $\lambda\in P_+$ for the $A$-series,     
\ie\ su($n)=A_{n-1}$, looks like $\la=\sum_{i=0}^{n-1}\la_i\L_i$ where 
$\sum \la_i=k$. Charge conjugation $\omega=C$ for su$(n)$ 
takes the weight $\la$ 
to the weight $C\la=\la_0\L_0+ \sum_{i=1}^{n-1}\la_{n-i}\L_i$. 
Charge conjugation is non-trivial for su($n$), except for su(2).

The exponents $\E$ here are the $C$-invariant weights, \ie\ those
$\mu$ with Dynkin labels $\mu_{n-i}=\mu_i$ for $1\le i\le n-1$. All of
these appear with multiplicity 1. The eigenvalues, involving ratios of
$S$-matrix elements, can be effectively computed as follows.

To the weight $\la=\sum_i\la_i\L_i$ associate orthogonal
coordinates $\la^+[\ell]=n-j+\sum_{i=\ell}^{n-1}\la_i$ for 
$\ell=1,\ldots,n$,\footnote{$^\star$}{We shall always use the
convention that $\sum_{i=r}^{r-1}=0$.} and write 
$t^+(\la)={n\,(n-1)\over 2}+\sum_{\ell=1}^{n-1}\ell\la_\ell$. Then 
\eqn\Ssun{
S_{\la\mu}=s\exp\left[2\pi\i\,{t^+(\la)\,t^+(\mu)\over (k+n)\,n}\right]\, 
\,{\rm det}\left(
\exp\left[-2\pi\i\,{\la^+[i]\,\mu^+[j]\over k+n}\right]
\right)_{1\le i,j\le n} \,,}
where $s$ is some (for our purposes irrelevant) constant that is given
in {\it e.g.} \refs{\gannon}. Here `det' denotes the determinant of the 
$n\times n$ matrix whose $(i,j)$-th entry is provided.

In order to describe the formula for $\psi$, we need to distinguish
two cases depending on the parity of $n$.
\medskip

\noindent {\bf The case of su($2n+1$)}

\noindent As we know, the exponents, which parametrise the
eigenvectors, and label the columns of $\psi$, are the $C$-invariant
weights of $\g=\widehat{{\rm su}}(2n+1)$ level $k$. Alternatively, we
can think of them as the level $k$ weights of the orbit Lie algebra
$\check{\g}=A_{2n}^{(2)}$. In either case, they can be equated with
the ($n$+1)-tuples $\mu=(\mu_0;\mu_1,\ldots,\mu_n)$, where  
$\mu_i\in \Zop_{\ge 0}$ and we have
$k=\mu_0+2\mu_1+\cdots+2\mu_{n-1}+2\mu_n$. To each exponent $\mu\in
\E$ we associate the coordinates $\mu[i]=n+1-i+\sum_{j=i}^n \mu_j$
for $i=1,\ldots,n$. 

The boundary states, which label the components of each eigenvector,
and both the rows and columns of each NIM-rep matrix 
$\N_\la=(\N_{\la a}^b)$, as well as the rows of $\psi$, 
are the level $k$ $C$-twisted weights of 
$\g=\widehat{{\rm su}}(2n+1$), or alternatively the level $k$ weights
of the twisted Lie algebra $\g^\omega=A_{2n}^{(2)}$. They can be
equated with all ($n+1$)-tuples $(a_0;a_1,\ldots,a_n)$ where 
$k=a_0+2a_1+2a_2+\cdots+2a_n$ and $a_i\in \Zop_{\ge 0}$. To each such
$a$ we associate the coordinates 
$a[i]=n+1-i+\sum_{j=i}^{n} a_j$ for $i=1,\ldots,n$.

\noindent Then the $\psi$-matrix, that is to say the modular matrix
$\hat{S}$ for $A_{2n}^{(2)}$, is\footnote{$^\dagger$}{This determinant
formula for $\hat{S}$ for the case of $A_{2n}^{(2)}$ first appeared in 
\refs{\fss}. They also noted that when $k$ is odd, $\hat{S}$ equals
the $S$-matrix for $\widehat{C}_n$ level ${k-1\over 2}$.} 
\eqn\psisuodd{
\psi_{a\mu}=(-1)^{{n(n-1)\over 2}} 
{2^n\over (k+2n+1)^{{n\over 2}}} \;
{\rm det}\left[
\sin\left({2\pi \,a[i]\,\mu[j]\over k+2n+1}\right)
\right]_{1\le i,j\le n}\,.}
We will show in section~5 that the column of $\psi$ corresponding to
$\mu^0=(k;0,\ldots,0)$, and the row of $\psi$ corresponding to 
$a^0=(k;0,\ldots,0)$ are both strictly positive. 

Incidentally, this $\psi$-matrix can be regarded as a symmetric
submatrix of the $S$-matrix for $\widehat{B}_n$ level $k+2$, provided
we identify $a$ and $\mu$ with the appropriate weights 
$a',\mu'$ in $P_+^{k+2}(B_n)$. More specifically, we identify 
$\mu\mapsto \mu'=(\mu_0+\mu_1+1;\mu_1,\ldots,\mu_{n-1},2\mu_n+1)$
and $a\mapsto
a'=(a_0+a_1+1;a_1,\ldots,a_{n-1},2a_n+1)$.\footnote{$^\ddagger$}{
$P_+^{k}(B_n)$ 
consists of the weights $\mu'=(\mu_0';\mu_1',\ldots,\mu_n')$ for which 
 $\mu'_0+\mu'_1+2\sum_{i=2}^{n-1} \mu'_i+\mu'_n=k$.}
With these identifications, the $\psi$-matrix becomes symmetric, and  
the relation between $\psi$ and the $\widehat{B}_n$ level $k+2$
$S$-matrix $S'$ is simply $\psi_{a\mu}=2\,S'_{a'\mu'}$. 

Using this identification, we can express the NIM-rep coefficients
$\N_{\la a}^b$, corresponding via \twisNIM\ to \psisuodd, in terms of
the ordinary fusion coefficients $N'$ of $\widehat{B}_n$ level $k+2$,
and branching rules for $B_n\subset {\rm su}(2n+1)$,
\eqn\nimsuodd{
\N_{\la a}^b=\sum_{\ga'}b^\la_{\ga'}\,\bigl(N'_{\ga',a'}
{}^{b'}-N'_{J\ga',a'}{}^{b'}\bigr)\,.}
Here, $J'$ is the simple current of $\widehat{B}_n$, which acts on 
weights $\nu'$ by $J\nu'=(\nu'_1;\nu'_0,\nu'_2,\ldots,\nu'_n)$. The
coefficients $b^\la_{\ga'}$ describe the branching rules 
$\oplus_{\ga'}\,b^\la_{\ga'}\,(\ga')=(\la)$ for the embedding of
$B_n\subset {\rm su}(2n+1)$; some branching
rules for $B_n\subset {\rm su}(2n+1)$ are given for instance in 
\refs{\mp}. 
The sum in \nimsuodd\ is over all dominant weights $\ga'$ of $B_n$;
we extend the definition of the fusion coefficient
$N'_{\ga',a'}{}^{b'}$ to arbitrary dominant weight $\ga'$ in the
obvious way (compare the Kac-Walton fusion formula): the fusion
coefficient is ${\rm det}(w)\, N'_{w.\ga',a'}{}^{b'}$ if there is a
unique element $w$ of the affine Weyl group of $\widehat{B}_n$ for
which $w.\ga':=w(\ga'+\rho)-\rho$ lies in $P_+^{k+2}(B_n)$, and it is
zero if there are more than one such $w$'s (\ie\ if $\ga'$ lies on the  
boundary of the Weyl alcove). Equation \nimsuodd\ makes it manifest
that our coefficients  $\N_{\la a}^b$ are integers.   
\smallskip

When the level $k$ is odd, a simplification occurs in that we can 
identify the boundary state $a$ with the entire $\widehat{C}_n$ level 
${k-1\over 2}$ weight 
$\widetilde{a}=({a_0-1\over 2};a_{1},\ldots,a_n)$, and similarly for  
$\mu$. With this, $\psi$ becomes the $\widehat{C}_n$ level ${k-1\over 2}$
$S$-matrix, and the NIM-rep becomes  
\eqn\nimsuodkev{
\N_{\la a}^b=\sum_{\tilde{\ga}}b_{\tilde{\ga}}^\la\,
\widetilde{N}_{\tilde{\ga}\tilde{a}}{}^{\tilde{b}}\,,}
where $\oplus \,b_{\tilde{\ga}}^\la\,(\tilde{\ga})=(\la)$ are the
branching rules for $C_n\subset {\rm su}(2n+1)$. As for \nimsuodd, the
fusion coefficients $\widetilde{N}_{\tilde{\ga}\tilde{a}}{}^{\tilde{b}}$
are defined for $\tilde{\ga}$ not necessarily in $\tilde{P}_+$, in
which case they can be negative. Some branching rules for 
$C_n\subset {\rm su}(2n+1)$ are given in \refs{\mp}.

This simplification reveals a hidden symmetry in the NIM-reps at odd
level $k$: the $\widehat{C}_n$ simple current $\tilde{J}=:J^B$ acts on
$a\in\B\cong P_+^{{k-1\over 2}}(C_n)$ by
${J}^Ba=(2a_n+1;a_{n-1},\ldots,a_1,{a_0-1\over 2})$ 
with phase $Q^B(\mu)={1\over 2}\sum_{i=1}^n i\mu_i$, 
and $\N_{\la,J^Ba}^{J^Bb}=\N_{\la a}^b$ (see \scnim).

Some low-rank clarifications are needed in both \nimsuodd\ and
\nimsuodkev. By $\widehat{B}_1$ level $k+2$ we mean affine su(2) at
level $2(k+2)$. Also, su(2)$=$so(3) is embedded in su(3) in two
different ways; one has branching rule $(2)=(1,0)$ and the other has
$(0)\oplus(1)=(1,0)$. By  $B_1\subset {\rm su}(3)$ we mean the former
embedding, and by $C_1\subset {\rm  su}(3)$ we mean the
latter. Similarly sp(4)$=$so(5) is embedded in su(5) in two different
ways: one with branching rule $(1,0)=(1,0,0,0)$ (using the so(5)
Dynkin labelling) and the other with $(1,0)\oplus(0,0)=(1,0,0,0)$
(using the sp(4) Dynkin labelling). By $B_2\subset {\rm su}(5)$ we
mean the former and by $C_2\subset{\rm  su}(5)$ we mean the latter. 
\medskip

\noindent {\bf The case of su($2n$)}

\noindent Next we determine the formula for $\psi$ for
$\g=\widehat{{\rm su}}(2n)$ at level $k$. The twisted algebra
$\g^\omega$ here is $\g^\omega=A_{2n-1}^{(2)}$ (for $n>2$), or $D_3^{(2)}$
with nodes 1 and 2 interchanged (for $n=2$). The rows of $\psi$  
are now labelled by $(n+1)$-tuples $a$ with  
$k=a_0+a_1+2a_2+\cdots+2a_n$.
To each such $a$ we associate the coordinates
$a[i]=n+1-i+\sum_{j=i}^{n} a_j$ for
$i=1,\ldots,n$. The orbit Lie algebra is $\check{\g}=D_{n+1}^{(2)}$;
 each exponent $\mu\in \E$ can be regarded
as an ($n$+1)-tuple $\mu$, where $k=\mu_0+2\mu_1+\cdots+2\mu_{n-1}+\mu_n$.  
To it we associate the coordinates 
$\mu[i]=2n+1-2i+2\sum_{j=i}^{n-1}\mu_j+\mu_n$.
Then the $\psi$-matrix (\ie\ $\widehat{S}$ for $A_{2n-1}^{(2)}$) is
\eqn\sueven{
\psi_{a\mu} = (-1)^{{n^2-n\over 2}}\, 
(k+2n)^{-{n\over 2}}\, 2^{n-\half} \; 
{\rm det}\left[
\sin\left({\pi\, a[i]\, \mu[j]\over k+2n}\right)
\right]_{1\leq i,j \leq n}\,.}
Although $\psi$ is non-symmetric, both the column corresponding
to $\mu^0=(k;0,\ldots,0)$ and the row corresponding to 
$a^0=(k;0,\ldots,0)$ are strictly positive.

We can interpret the resulting NIM-rep coefficients 
$\N_{\la a}^b$ using fusions $N'$ and branching coefficients
$b_{\ga'}^\la$ for the semisimple algebra 
${\rm su}(2)\oplus\cdots\oplus{\rm su}(2)$ ($n$ times) at level
$k+2n-2$  
\eqn\nimsuev{
\N_{\la a}^b=\sum_{d}\sum_{\ga'}\sum_\pi
\epsilon(\pi)\,b_{\ga'}^\la\,N'_{J^d\ga',\pi a'}{}^{b'}\,.}
This formula will be explained in more detail in section~5. We can
replace the sum over $\pi$ with the determinant of an $n\times n$
matrix made up of various $A_1$ fusions. Even so, equation \nimsuev\
is not very practical, but it does make it manifest that our
coefficients $\N_{\la a}^b$ are integers.  

The symmetry of the Dynkin diagram of $\check{\g}=D_{n+1}^{(2)}$ 
gives rise to a simple current symmetry $J^E$ of $\E$, while the
symmetry of the Dynkin diagram of $\g^\omega=A_{2n-1}^{(2)}$ yields a
simple current symmetry $J^B$ of $\B$ --- see \scnim. These will be
described in section~5.3.

\subsec{The D-series}

\noindent Next we consider the algebra 
$\widehat{{\rm so}}(2n)=\widehat{D}_n$ 
at level $k$. A weight $\la \in P_+$ here satisfies 
$k=\la_0+\la_1+\la_{n-1}+\la_n+2\sum_{i=2}^{n-2}\la_i$.   

As before, we need to compute effectively certain ratios of $S$-matrix
elements. In order to do so  we associate to the weight
$\la=(\la_0;\la_1,\la_2,\ldots,\la_n)$ the orthogonal coordinates  
$\la^+[\ell]=n-\ell+\sum_{i=\ell}^{n-1}\la_i+{\la_n-\la_{n-1}\over 2}$ 
for $\ell=1,\ldots,n$. Then
\eqn\Sson{
S_{\la\mu}=s\left\{{\rm det}
\left[\cos\left(2\pi\,{\la^+[i]\,\mu^+[j]\over k+2n-2}\right)
\right]_{1\le i,j\le n} + (-\i)^n \,
{\rm det}\left[\sin\left(2\pi\,{\la^+[i]\,\mu^+[j]\over k+2n-2}\right)
\right]_{1\le i,j\le n} \right\}\,,}
where $s$ is some irrelevant constant (which is again given in
{\it e.g.} \refs{\gannon}). 
\medskip

\noindent {\bf The case of so$(2n)$ with chirality flip}

\noindent The chirality flip $\omega=P$ interchanges the $n$th and 
($n-1$)st Dynkin labels, $\la_n\leftrightarrow \la_{n-1}$. The
chirality flip $P$ agrees with charge conjugation $C=S^2$ when $n$ is
odd, but $C$ is trivial when $n$ is even. The relevant orbit Lie
algebra here is $\check{\g}=A_{2n-3}^{(2)}$.  
The exponents are the $P$-invariant $\widehat{D}_n$ weights, \ie\
the weights of the form
$(\mu_0;\mu_1,\mu_2,\ldots,\mu_{n-1},\mu_{n-1})$ 
where $k=\mu_0+\mu_1+2\mu_2+\cdots+2\mu_{n-1}$. 
Given any $P$-invariant $D_n^{(1)}$ weight $\mu$, define its
coordinates by $\mu[i]=n-i+\sum_{j=i}^{n-1}\mu_j$, for
$i=1,\ldots,n-1$. 

The relevant twisted algebra here is $\g^\omega=D_{n}^{(2)}$. The
boundary states are all $n$-tuples $(a_0;a_1,\ldots,a_{n-1})$ of 
non-negative integers such that 
$k=a_0+2a_1+2a_2+\cdots+2a_{n-2}+a_{n-1}$.  
Define coordinates 
$a[i]=2n-1-2i+2\sum_{j=i}^{n-2}a_j+a_{n-1}$ for $i=1,\ldots,n-1$.
Then the $\psi$-matrix (\ie\ the matrix $\widehat{S}$ of
$D_{n}^{(2)}$) is given by 
\eqn\son{
\psi_{a\mu}=\left({1\over k+2n-2}\right)^{{n-1\over 2}}\,
2^{n-2}\,\sqrt{2}\; \i^{n^2+n+2}\, 
{\rm det}\left[\sin\left({\pi\, a[i]\,\mu[j]\over k+2n-2}\right)
\right]_{1\le i,j\le n-1}\,.}
Again $\psi$ is non-symmetric, but both its $\mu^0=(k;0,\ldots,0)$ 
column and $a^0=(k;0,\ldots,0)$ row are strictly positive. 

The NIM-rep coefficients can be interpreted in terms of fusions $N'$
of $\widehat{B}_{n-1}$ at level $k+1$, and branching coefficients
$b_{\ga'}^\la$ of $B_{n-1}\subset$ so$(2n)$,
\eqn\nimso{
\N_{\la a}^b=\sum_{\ga'}
b_{\ga'}^\la\,\bigl(N'_{\ga',a'}{}^{b'}-N'_{J\ga',a'}{}^{b'}\bigr)\,.}
This will be explained in section~5.2; as before it demonstrates that 
our coefficients $\N^b_{\la a}$ are integers.

The symmetry of the Dynkin diagram of $\check{\g}=A_{2n-1}^{(2)}$ 
gives rise to a simple current symmetry $J^E$ of $\E$, while the
symmetry of the Dynkin diagram of $\g^\omega=D_{n}^{(2)}$ yields a
simple current symmetry $J^B$ of $\B$ --- see \scnim. These will be
described in section~5.3.

\medskip

\noindent {\bf The case of so$(8)$ with triality}

\noindent There are 4 other non-trivial conjugations for so(8), but
their NIM-reps are determined (see the end of this subsection) once we
know the NIM-rep corresponding to `triality' $\omega=T$.  
The order-$3$ triality maps the Dynkin
labels $\mu=(\mu_0;\mu_1,\mu_2,\mu_3,\mu_4)$  to
$(\mu_0;\mu_4,\mu_2,\mu_1,\mu_3)$. The relevant twisted algebra here 
is $\check{\g}=\g^\omega=D_4^{(3)}$. The exponents are the
$\widehat{D}_4$ weights of the form  $(\mu_0;\mu_1,\mu_2,\mu_1,\mu_1)$
where $k=\mu_0+3\mu_1+2\mu_2$. Given any such exponent $\mu$, define 
coordinates $\mu[1]=3\mu_1+\mu_2+4$,
$\mu[2]=\mu_2+1$. 

The boundary states are labelled by  all triples $(a_0;a_1,a_2)$ of
non-negative integers such that $k=a_0+ 2a_1+3a_2$. Define
coordinates  $a[1]=a_1+a_2+2$, $a[2]=a_2+1$. Put $\kappa=k+6$. Then
the $\psi$-matrix (\ie\ the matrix $\widehat{S}$ for $D_4^{(3)}$) is 
\eqn\gtwoS{\eqalign{
\psi_{a \mu}=& {2\over \kappa}\, \Bigl( c(m m'+2 m n'+2 n m'+ n
n')+c(2 m m'+m n'+n m'-n n')\cr &\qquad +c(-m m'+m n'+n m'+2 n n')
-c(2 m m'+m n'+n m'+2 n n')\cr &\qquad -c(m m'+2 m n'-n m'+n n') 
-c(m m'-m n'+2 n m'+n n') \Bigr)\,,}}
where $c(x)=\cos({2\pi x \over 3\kappa})$, and we take 
$m=a[1]$, $n=a[2]$, $m'=\mu[1]$,
$n'=\mu[2]$.
Again $\psi$ is non-symmetric, but both its $\mu^0=(k;0,0,0,0)$ 
column and $a^0=(k;0,0,0,0)$ row are strictly positive. 

We can find an interpretation of the NIM-rep coefficients $\N$ in
terms of fusions $N'$ and branching coefficients $b_{\ga'}^\la$ of
$A_2$ at level $k+3$
\eqn\dfournim{
\N_{\la a}^b=\sum_{i=0}^2\sum_{\ga'}b_{\ga'}^\la\,
\Bigl(N'_{J'{}^i\ga',a'}{}^{b'}-
N'_{J'{}^i\ga',C'a'}{}^{b'}\Bigr)\,.}
This is further explained in section~5. The formula \dfournim\ 
implies, in particular, that our NIM-rep is manifestly integral.  

\smallskip We can now give the NIM-reps for the 3 remaining
conjugations of so(8). The NIM-rep for the conjugation  
$T^{-1}:\la_1\rightarrow \la_4\rightarrow \la_3\rightarrow\la_1$
equals that for triality $T$, given above. The NIM-rep for the
conjugation $T^{-1}P:\la_1\leftrightarrow\la_3$ is given by
$\la\mapsto \N^P_{T\la}$, where $\N^P$ is the NIM-rep for chirality
flip, while that for $TP:\la_1\leftrightarrow\la_4$ is 
$\la\mapsto \N^P_{T^{-1}\la}$.

\subsec{The algebra $E_6$}

\noindent The final algebra with a non-trivial conjugation is
$E_6$. Its level $k$ weights $\la$ satisfy  
$k=\la_0+\la_1+2\la_2+3\la_3+2\la_4+\la_5+2\la_6$. 
The order-2 charge conjugation $\omega=C$ interchanges the 1st and 
5th, and 2nd and 4th, Dynkin labels, and fixes the 0th, 3rd, and
6th. The relevant twisted algebra here is
$\check{\g}=\g^\omega=E_6^{(2)}$.  

The ratios of $S$-matrix elements can be effectively calculated using
the formula of \refs{\gannon}.
The exponents are the $C$-invariant $E_6$ weights, i.e.\ the $\mu$ of
the form $(\mu_0;\mu_1,\mu_2,\mu_3,\mu_2,\mu_1,\mu_6)$ where
$k=\mu_0+2\mu_1+4\mu_2+3\mu_3+2\mu_6$. Define   coordinates by
$\mu[1]=2\mu_1+3\mu_2+2\mu_3+\mu_6+8$,
$\mu[2]=\mu_2+\mu_3+\mu_6+3$, $\mu[3]=\mu_2+\mu_3+2$,
and $\mu[4]=\mu_2+1$. 

The boundary states are all quintuples $(a_0;a_1,a_2,a_3,a_4)$ of
non-negative integers such that $k=a_0+ 2a_1+3a_2+4a_3+2a_4$.  
Define coordinates 
$a[1]=a_1+{3\over 2} a_2+2a_3+a_4+ {11\over2}$,
$a[2]={1\over 2} a_2+a_3+a_4+{5\over 2}$, 
$a[3]={1\over 2} a_2+a_3+{3\over 2}$, and 
$a[4]={1\over 2} a_2+{1\over 2}$.
Put $\kappa=k+12$. Then the $\psi$-matrix (\ie\ the matrix
$\widehat{S}$ for $E_6^{(2)}$) is 
\eqn\esix{\eqalign{
\psi_{a\mu}=&\,{16\over \kappa^2} \Biggl\{
{\rm det}\left[\sin\left({2\pi\, a[i]\,\mu[j]\over \kappa}\right)
\right]_{1\le i,j\le 4}
+{\rm det}\left[\sin\left({2\pi\, (c.a)[i]\,\mu[j]\over \kappa}\right)
\right]_{1\le i,j\le 4}\cr 
&\qquad +{\rm det}\left[\sin\left({2\pi\, (c^t.a)[i]\,\mu[j]\over
\kappa}\right) \right]_{1\le i,j\le 4} \Biggr\}\,.}}
Here, $(c.a)[i]$ denotes the $i$th coordinate of
the matrix product of $c$ with the column vector with entries
$a[j]$, and $c^t$ means the transpose(=inverse) of $c$, where
$c$ is the orthogonal matrix 
\eqn\cee{c={1\over 2}\left(\matrix{1&1&1&-1\cr 1&1&-1&1\cr 
1&-1&1&1\cr 1&-1&-1&-1}\right)\,.}
Again, $\psi$ is non-symmetric but the $a^0=(k;0,\ldots,0)$ row is 
strictly positive, and the same appears to hold for the 
$\mu^0=(k;0,\ldots,0)$ column. The positivity of the $a^0$ row can be 
shown in the usual manner, but the arguments to prove the positivity
of the $\mu^0$ column are much messier; we have therefore only
verified this claim on a computer for levels up to $k=100$. 

We can  interpret our
$\widehat{E}_6$ NIM-reps using $\widehat{{\rm so}}(8)$ fusion
coefficients $N'$ and branching coefficients $b_{\ga'}^\la$
\eqn\nimesix{\N_{\la a}^b=\sum_{J'}\sum_{\ga'}
\sum_{\pi}b_{\ga'}^\la\,\epsilon(\pi)\, N'_{J'\ga',\pi a'}{}^{b'}\,.} 
This sum \nimesix\ is manifestly an integer.

\newsec{Calculations and further elaborations}

In this section we want to provide details of the derivation of
the various formulae we gave in the previous section. First we shall
explain how we obtained the formulae for $\hat{S}=\psi$. As an example
we shall discuss the calculation for the algebra 
$\bar{\g}={\rm su}(2n)$ explicitly. 

\subsec{Derivation of the $\psi$-formula for 
$\bar{\g}={\rm su}(2n)$} 

{}From the discussion of section~3, we need to identify the
quantities of Thm.~13.9 in \refs{\kac}. The finite Weyl group of both 
$\check{\g}=D_{n+1}^{(2)}$ and $\g^\omega=A_{2n-1}^{(2)}$ is just the
Weyl group of sp$(2n)$, so the $\psi$ formula will involve a
determinant that will look like the one in the $S$-matrix for
$\widehat{{\rm sp}}(2n)$. Furthermore, we need to be able to evaluate
inner products $(a|\mu)$, or equivalently we need the values of
$(\L_i|\L_j')$ for the fundamental weights $\L_i,\L_j'$ of $\g^\omega$
and $\check{\g}$, respectively. To do this we use the equation (see
page 222 of \refs{\kac}) that $(\alpha_j^\vee|\L_i)=\delta_{ij}$ for
the coroots $\alpha_j^\vee$ of ${\g}^\omega$, where the values of
$(\alpha_i^\vee|\alpha_j^\vee)$ are determined from the Cartan matrix
of $A^{(2)}_{2n-1}$ (see (6.2.1) of \refs{\kac}). We obtain the
realisation $\L_i=e_1+\cdots+e_i$ for all $i=1,\ldots,n$, where 
$e_i$ is an orthonormal basis of $\Rop^n$. We now find $\L_i'=\L_i$
for $i<n$, and $\L_n'={1\over {2}}\L_n$, by page 266 of \refs{\kac}. 
This implies that we should define the orthogonal coordinates 
$\mu[i]$ and $a[i]$, as we did in subsection~4.1, and rescale the
inner products by 2.

The overall multiplicative factor is computed as follows. The number
of positive roots of the horizontal algebra $C_n$ of ${\g}^\omega$ is
$n^2$. The lattice $M'$ is the lattice spanned by the coroots
$\alpha_i^\vee$, which is the square lattice  $\Zop^n$. The lattice
$M$ is the lattice spanned by the roots $\alpha_i$, which is the
$D_n$ root lattice (see page 93 of \refs{\kac}). The index
$|M'/M|$ is $\sqrt{2}$, and the index $|M^*/(k+h^\vee)M|$ is
$4\,(k+2n)^n$. We want to replace $e^{-\i z}-e^{\i z}$ with
$-2\i\sin(z)$, and so we get an additional factor of 
$2^n \i^{-n}$. Collecting all these terms,  we recover \sueven.

\subsec{Relation of NIM-reps to ordinary fusion rules}

Now we turn to the derivations of the fusion formulae for our
NIM-reps. Consider first the easiest case: su$(2n+1)$ when $k$ is
odd. Here, we noticed that 
$\psi_{a\mu}=\widetilde{S}_{\tilde{a}\tilde{\mu}}$, 
where the tilde denotes quantities for $C_n$ level ${k-1\over 2}$. The
map $\mu\mapsto \tilde{\mu}$ described in section 4.1 is onto all of 
$P_+^{{k-1\over 2}}(C_n)$. We know that an su$(2n+1)$ character,
restricted to the Cartan subalgebra (CSA) of 
$C_n\subset {\rm su}(2n+1)$, equals the sum of $C_n$ characters, the
sum given by the appropriate branching coefficients
$b_{\tilde{\ga}}^\la$. Now, the su$(2n+1)$ ratio 
${S_{\la\mu}\over S_{0\mu}}$ is the $\la$-character of su$(2n+1)$ 
evaluated at the element $-2\pi\i\,{\mu+\rho\over k+2n+1}$ in the 
CSA not only of su$(2n+1)$, but in fact of the embedded algebra
$C_n$, since $C\mu=\mu$. This means that it equals the sum
$\sum_{\tilde{\ga}}b_{\tilde{\ga}}^\la\,\widetilde{S}_{\tilde{\ga}
\tilde{\mu}}/\widetilde{S}_{\tilde{0}\tilde{\mu}}$
of $C_n$ characters evaluated at that element of the CSA of $C_n$.
Thus we have succeeded in expressing all su$(2n+1)$ quantities in
\twisNIM\ in terms of $C_n$ quantities, and we obtain \nimsuodkev\
from \verlinde\ and 
\eqn\suevcalc{
\N_{\la a}^b=\sum_{\mu\in\E}{\psi^*_{b\mu}\,
S_{\la\mu}\,\psi_{a\mu}\over S_{0\mu}}=
\sum_{\tilde{\ga}}b_{\tilde{\ga}}^\la\,
\sum_{\tilde{\mu}\in\tilde{P}_+} 
{\tilde{S}^*_{\tilde{b} \tilde{\mu}}\,
\tilde{S}_{\tilde{\ga}\tilde{\mu}}\,
\tilde{S}_{\tilde{a}\tilde{\mu}}\over 
\tilde{S}_{\tilde{0}\tilde{\mu}}}\,,}
where here (as throughout this paper) we extend the definition of
fusion coefficients and $S$-entries to arbitrary dominant weights, as
explained after \nimsuodd.\medskip

Next we want to verify \nimsuodd. The only modification that arises
relative to the previous case is that the set of all $\mu'$ is only a
subset of all weights in $P_+':=P_+^{k+2}(B_n)$, namely all those
weights with $\mu_n'$ odd and $\mu'_0>\mu_1'$. In order to recover the
$B_n$ fusions from \twisNIM, we need to extend the sum from all these
$\mu'$ associated to $\mu\in\E$, to all weights $\mu'\in P_+'$. This
is where the simple current $J'$ comes in. Its phase $Q(\mu')$ is
${\mu'_n \over 2}$ (recall \simcur), and  so the alternating sum in
\nimsuodd\ projects away $\mu_n'$ even, as well as $\mu'_0=\mu'_1$,
and absorbs an extra  factor of 2 coming from our formula
$\psi=2S'$. The weights with $\mu'_0<\mu_1'$ are recovered by taking
$J'$-orbits of $\mu'$, and this absorbs the remaining factor of 2. We
have also used here that $Q(\ga')\in\Zop$ and 
$Q(a'),Q(b')\in \Zop+{1\over 2}$.  
\medskip

\noindent {\bf The case of su$(2n)$} 

\noindent In this case, the $\psi$-matrix is a submatrix of $S$
(rescaled by $\sqrt{2}$) for $\widehat{B}_n$ level $k+1$,
using the identifications  
$\mu\mapsto (\mu_0+\mu_1+1;\mu_1,\ldots,\mu_n)\in
P_+^{k+1}(B_n)$ and 
$a\mapsto (a_0;a_1,\ldots,a_{n-1},2a_n+1)\in
P_+^{k+1}(B_n)$. 
But in order to interpret the resulting NIM-rep coefficients 
$\N_{\la a}^b$ using ordinary fusions $N'$, we need to break the
algebra down further. 

Write $A_1^n$ for the direct sum 
${\rm su}(2)\oplus\cdots\oplus{\rm su}(2)$ ($n$ times), and put 
$P_+'=P_+^{k+2n-2}(A_1)$. Given an exponent $\mu\in \E$, define the
weight $\mu'=(\mu'{}^{(1)}, \ldots,\mu'{}^{(n)})\in P_+'$ by
$\mu'{}^{(i)} =(k+2n-2-\mu[i];\mu[i])\in
P_+^{k+2n-2}(A_1)$. Similarly, define $a'\in P_+'$ by 
$a'{}^{(i)}=(k+2n-2-a[i];a[i])\in P_+^{k+2n-2}(A_1)$. The quantities
$b_{\ga'}^\la$ in \nimsuev\ are the
$A_1^n\subset {\rm su}(2n)$ branching coefficients
$\oplus_{\ga'}\,b_{\ga'}^\la\,(\ga')=(\la)$.
The first sum in \nimsuev\ is over all $n$-tuples $d\in\Zop_2^n$ with
$\sum_i d_i$ even. $A_1$ has a simple current $J'$ which sends $\nu'$
to $(\nu'_1;\nu'_0)$, and by $J'{}^d\ga'$ we mean the weight with 
$(J'{}^d\ga')^{(i)}=J'{}^{d_i}(\ga'{}^{(i)})$. The third sum in 
\nimsuev\ is over all permutations $\pi\in S_n$, which act on $a'$ by
$(\pi a')^{(i)}=a'{}^{(\pi i)}$. The quantity $\epsilon(\pi)$ is the
sign of $\pi$. 

\noindent Expand out \sueven\ into 
\eqn\psian{
\psi_{a\mu}={2^{n-{1\over 2}}\over (k+2n)^{n\over 2}}\i^{n^2-n}\sum_\pi  
\epsilon(\pi)\, \prod_i\sin\Bigl({\pi\,a[\pi i]\,\mu\over k+2n}\Bigr) 
= 2^{n-1\over 2}\i^{n^2-n}\sum_\pi \epsilon(\pi)\,S'_{\pi a',\mu'}\,.}
The set of all $\mu'$ here consist precisely of all the weights in
$P_+'$ with $\mu'_1{}^{(1)}>\mu'_1{}^{(2)}>\cdots>\mu'_1{}^{(n)}$ and 
$\mu'_1{}^{(1)}\equiv \cdots\equiv \mu'_1{}^{(n)}$ (mod 2). Again, we
need to extend the sum over these $\mu'$ to a sum over all $P_+'$. The
non-trivial simple current of $A_1$ has phase $Q(\nu)={\nu_1\over 2}$. 
Thus the sum in \sueven\ over the $2^{n-1}$ $n$-tuples $d$ will retain
only those $\mu'$ with 
$\mu'_1{}^{(1)}\equiv \cdots\equiv \mu'_1{}^{(n)}$ (mod 2), and with
the numbers $\mu'_1{}^{(i)}$ all pairwise distinct. It also absorbs
the extra factor of $2^{n-1}$ coming from \psian. An additional sum
over the $S_n$-orbits of $\mu'$ removes the 
$\mu'_1{}^{(1)}>\mu'_1{}^{(2)}>\cdots>\mu'_1{}^{(n)}$
restriction. This also cancels an extra factor of $n!$ coming from our 
over-counting of fusion coefficients due to the sum over $S_n$-orbits
of both $a'$ and $b'$ --- we use here the fact that the branching
coefficients  $b_{\ga'}^\la$ and $b_{\pi\ga'}^\la$ are equal, for any
$\pi$. 

\medskip \noindent {\bf The case of so$(2n)$} 

\noindent Write $P_+'=P_+^{k+1}(B_{n-1})$. Under the identification 
$\mu\mapsto \mu'=(\mu_0;\mu_1,\ldots,\mu_{n-2},2\mu_{n-1}+1)\in
P_+'$ and
$a\mapsto a'=(a_0+a_1+1;a_1,\ldots,a_{n-1})\in P_+'$, the
$\psi$-matrix \son\ is seen to be a submatrix (up to a factor of
$\sqrt{2}$) of the $S$-matrix of  $\widehat{B}_{n-1}$ at level $k+1$. 

Recall that the simple current $J$ of $\widehat{B}_{n-1}$ acts on
$\ga'$ by interchanging $\ga'_0\leftrightarrow\ga_1'$, and has phase
$Q(\ga')={\ga'_{n-1}\over 2}$. The $\mu'$ consist of all weights in
$P_+'$ with $\mu'_{n-1}$ odd (\ie\ the spinors). The coefficients
$b_{\ga'}^\la$ in \nimso\ are the branching rules 
$\oplus_{\ga'\in P_+'}\,b_{\ga'}^\la(\ga')=(\la)$ for the embedding  
$B_{n-1}\subset D_n$. Some simple branching rules are
$(\L_1')+(0')=(\L_1)$, $(\L_i')+(\L_{i-1}')=(\L_i)$ for $1<i<n-1$, and
$(\L'_{n-1})=(\L_{n-1})= (\L_n)$; others can be found for instance 
in \refs{\mp}.

\medskip

\noindent {\bf The case of so$(8)$} 

\noindent Using the identifications 
$\mu\mapsto (\mu_0;\mu_2,3\mu_1+2)$ and
$a\mapsto(a_0+a_1+a_2+2;a_2,a_1)$, the $\psi$-matrix can be seen to be a  
submatrix of the level $k+2$ $S$-matrix of
$\widehat{G}_2$ (rescaled by $\sqrt{3}$).
It is clear from this interpretation of $\psi$ that the $a^0$ row of 
$\psi$ will be strictly positive. That the $\mu^0$ column, 
corresponding to the $\widehat{G}_2$ weight $(k;0,2)$, of $\psi$ is
strictly positive, is a little more awkward: we need to show
\eqn\gtwa{
c(2m+3n)+c(3m+n)+c(m-2n)>c(3m+2n)+c(2m-n)+c(m+3n)\,,} 
for all $m=a_1+a_2+2,n=a_2+1$, where here 
$c(x)=\cos({2\pi x\over \kappa})$; to do this, reduce it first to the
quadratic inequality 
\eqn\gtwb{
c(\delta)^2-(2c(\sigma)^3-c(\sigma))\,c(\delta)+(4c(\sigma)^4
-3c(\sigma)^2)\ge 0\,,} 
where $\sigma={m+n\over 2}={a_1+2a_2+3\over 2}$ and
$\delta={m-n\over 2}={a_1+1\over 2}$. It can be shown by standard
arguments that this final inequality is indeed satisfied 
for all $a\in \B$.

To get an interpretation of the NIM-rep coefficients $\N$ in terms of
ordinary fusions $N'$, we must break $G_2$ down to $A_2$. 
In particular, write $P_+'=P_+^{k+3}(A_2)$. Associate the exponent 
$\mu\in P_+^k(D_4)$ to the weight
$\mu'=(\mu_0;\mu_2,3\mu_1+\mu_2+3)\in P_+'$, and  
the boundary state $a$ to the weight
$a'=(a_0+a_1+a_2+2;a_2,a_1+a_2+1)\in P_+'$. In \dfournim,
$J'$ is the simple current of $A_2$, which sends $\nu'$ to
$(\nu'_2;\nu_0',\nu'_1)$. $C'$ there is charge conjugation, which
interchanges $\nu'_1\leftrightarrow\nu'_2$. The quantities 
$b_{\ga'}^\la$ are the $A_2\subset D_4$ branching coefficients
$\oplus_{\ga'}\,b_{\ga'}^\la\,(\ga')=(\la)$. There are two of these branchings;
the one we need comes from $A_2 \subset G_2 \subset B_3 \subset D_4$. 
Many of these
branching rules can be found in \refs{\mp}; some of the more useful ones
are $2(0,0)\oplus(1,0)\oplus(0,1) =(1,0,0,0)=(0,0,1,0)=(0,0,0,1)$ and 
$(1,1)\oplus 3(1,0)\oplus 3(0,1)\oplus 2 (0,0)=(0,1,0,0)$.  

\noindent Then \gtwoS\ becomes 
\eqn\dforato{
\psi_{a \mu}=\sqrt{3}\i\Bigl( S'_{a'\mu'}-S'_{a'\mu'}{}^*\Bigr)\,,} 
expressing $\psi$ in terms of $A_2$ data.

The $\mu'$  consist of all weights in $P_+'$ with
$\mu'_1<\mu_2'$ and $\mu'_1\equiv \mu_2'$ (mod 3). The first condition 
on $\mu'$ is removed by the sum of $C'$-orbits of $\mu'$ --- we use
the fact that the branching coefficients $b_{\ga'}^\la$ and
$b_{C'\ga'}^\la$ are equal. The second condition on $\mu'$ is removed
by the sum over $i$ in \dfournim\ (recall that the phase for $J'$ is
$Q(\nu')={\nu'_1-\nu'_2\over 3}$).  
\medskip 

\noindent {\bf The case of $E_6$}

\noindent In this case the $\psi$-matrix is a submatrix of the level 
$k+3$ $S$-matrix (rescaled by 2) of $\widehat{F}_4$, using the
identifications $\mu\mapsto(\mu_0;\mu_6,\mu_3,2\mu_2+1,2\mu_1+1)$ and 
$a\mapsto (a_0+a_3+a_2+a_1+3; a_4,a_3,a_2,a_1)$.
Now, any $\widehat{F}_4$ level $k$ quantities can be rewritten in
terms of $\widehat{{\rm so}}(8)$ level $k+3$ ones, using the embedding 
so(8)$\,\subset F_4$, and this will enable us to interpret these
NIM-reps using $\widehat{{\rm so}}(8)$ fusion coefficients.  

In particular, write $P_+'=P_+^{k+6}(D_4)$, and associate the exponent
$\mu\in P_+^k(E_6)$ to the weight 
$\mu'=(\mu_0;2\mu_1+2\mu_2+\mu_3+4,\mu_6,\mu_3,2\mu_2+\mu_3+2)\in
P_+'$ and the boundary state $a$ to
$a'=(a_0+a_1+a_2+a_3+3;a_1+a_2+a_3+2,a_4,a_3,a_2+a_3+1)\in
P_+'$. The sum in \nimesix\ over $J'$ is over the 4 simple currents of
so(8): the identity; $J'_v\nu'=(\nu'_1;\nu'_0,\nu'_2,\nu'_4,\nu'_3)$; 
$J'_s\nu'=(\nu'_4;\nu'_3,\nu'_2,\nu'_1,\nu'_0)$; and $J'_v\circ J'_s$. 
The sum there over $\pi$ is over all 6 conjugations of so(8),
corresponding to each of the 6 possible permutations $\pi$ of the 1st,
3rd, and 4th Dynkin labels, and $\epsilon(\pi)$ is the sign of that
permutation. The numbers $b_{\ga'}^\la$ are the $D_4\subset E_6$
branching coefficients
$\oplus_{\ga'}\,b_{\ga'}^\la\,(\ga')=(\la)$. Some useful 
$D_4\subset E_6$ branching rules are
$3(0000)\oplus(1000)\oplus(0010)\oplus(0001)=(100000)=(000010)$ and
$2(0000)\oplus 2(1000)\oplus(0100)\oplus 2(0010)\oplus 2(0001)=(000001)$. 

To prove \nimesix, note first that 
$\psi_{a\mu}=2\sum_\pi \epsilon(\pi)\,S'_{\pi a',\mu'}$, where the sum
is over all 6 permutations $\pi$ of the labels $a'_1,a'_3,a_4'$. The
weights $\mu'$ consist of all weights in $P_+'$ with
$\mu'_1>\mu_3'>\mu'_4$ and $\mu'_1\equiv\mu_3'\equiv\mu'_4$ (mod 2). 
The phases for the simple currents $J'_v$ and $J'_s$ are
$Q_v(\nu')={\nu'_3+\nu'_4\over 2}$ and 
$Q_s(\nu')={\nu_1'+ \nu'_3\over 2}$. The rest is as before. 

\subsec{Further remarks} 

It remains to show that, for the case of the $A$- and $D$-series, the
column of $\psi$ corresponding to $\mu^0=(k;0,\ldots,0)$, and the row 
of $\psi$ corresponding to  $a^0=(k;0,\ldots,0)$ are both strictly 
positive. 
The key observation for the proof of these statements is the fact that 
$\psi$ is a submatrix of some untwisted $S$-matrix (see
section~5.2). In some cases ({\it e.g.} for the $\mu^0$ column for 
the case of su($2n$)) strict positivity then follows automatically
from the fact that the vacuum row or column of that $S$-matrix is
strictly positive (see Remark 13.8 in \refs{\kac}). However, in most
other cases we need to know the sign of the $\L_n$ row and column of
the $S$-matrix for $\widehat{B}_n$.   

Consider for concreteness the $\mu^0=(k;0,\ldots,0)$ column of $\psi$, 
for su$(2n+1)$. We know from section~4.1 that 
$\psi_{a\mu^0}=2S'_{a'\L'_n}$. The sign of $S'_{a'\L_n'}$ equals the 
sign of ${S'_{\L'_na'}/ S'_{0'a'}}$, which is a value of the
$\L_n$-character of $B_n$. This is easy to calculate, and we find 
\eqn\spinor{
{S'_{\L'_n a'}\over S'_{0' a'}}=2^n\,
\prod_{i=1}^n\sin\Bigl({\pi\,a[i]\over k+2n+1}\Bigr)\,.}
Given the definition of $a[i]$, we therefore conclude that 
$\psi_{a' \mu^0}>0$. All other algebras are handled in  
the same way.  
\medskip 

Finally, let us discuss the simple current symmetries for the various
NIM-reps. The simple current $J^E$ for su$(2n$), which corresponds to 
the Dynkin symmetry of $D_n^{(2)}$, permutes $\mu\in\E$ by 
$J^E\mu=(\mu_n;\mu_{n-1},\ldots,\mu_0,\mu_1,\ldots,\mu_{n-1})$. Its
phase $Q^E(a)={1\over 2}\sum_iia_i$ will be a grading for the NIM-rep
coefficients (see \scnim), where 
$Q^E(\la)={1\over 2}\sum_{i=1}^{2n-1}i\la_i$. The simple current $J^B$
for su$(2n)$, which corresponds to the Dynkin symmetry of
$A_{2n-1}^{(2)}$, permutes $a\in\B$ by $a_0\leftrightarrow a_1$, and
has phase $Q^B(\mu)=\mu_n$. It defines a symmetry of the NIM-rep (see
\scnim). 

The simple currents $J^E$ and $J^B$ for so$(2n$) are defined
similarly, except with their formulae interchanged (for example $J^E$ 
acts on $\mu$ by interchanging $\mu_0\leftrightarrow\mu_1$, and we
have $Q^E(a)=a_n$). The NIM-rep grading \scnim\ arises with the choice 
$Q^E(\la)={1\over 2}(\la_{n-1}+\la_n)$.

\newsec{Generalisations}

\noindent The above constructions generalise also to certain other
classes of modular invariants, in particular those that come from
simple currents \simcur. Here we shall only give the
details for two simple cases; a complete description will be given
elsewhere. Together, these two cases cover all su($p$) level $k$,
where $p\ge 3$ is prime. The $\psi$-matrices for all modular
invariants of su(2) were given in \refs{\bppz} ($\psi$ for the
`tadpole' su(2) NIM-rep, which does not correspond to a modular 
invariant, is a special case of the next subsection). 
The simple current NIM-reps given below are presumably consistent
with those in \refs{\fhsw}, but we are more explicit here. We 
should also like to emphasise that, unlike us, \refs{\fhsw} do not
have a rigorous proof that their conjectured NIM-reps  are non-negative
integers (see the discussion in the introduction).  

Consider any su$(n)$ level $k$. It has a simple current $J$ of order
$n$, corresponding to the cyclic symmetry of the extended Dynkin
diagram, which sends the highest weight 
$\la=(\la_0;\la_1,\ldots,\la_{n-1})$ to
$J\la=(\la_{n-1}; \la_0,\ldots,\la_{n-2})$. For any divisor $d$ of
$n$, define the matrix $M[d]$ by 
\eqn\simcurM{
M[d]_{\la,\mu}=
\sum_{j=1}^{n/d}\delta^{{n\over d}}
\left(t(\la)+{d\, j\, k' \over 2}\right)\,\delta_{\mu,J^{jd}\la}\,,} 
where $t(\la)=\sum_ii\la_i$, and $k'=k+n$ or $k'=k$ depending on
whether or not both $k$ and $n$ are odd, respectively. In \simcurM\ we
write $\delta^y(x)=0$ or $1$ depending, respectively, on whether or
not ${x\over y}\in{\Bbb Z}$. Then $M[d]$ is a modular invariant if and
only if the product $(n-1)\, k\, d$ is even. For instance, $M[n]=I$ is
a modular invariant for any su($n$) level $k$; for su(2), $M[1]$ is a
modular invariant if and only if $k$ is even.

\subsec{A case without fixed points}

Let us consider su($n$) level $k$, for the simple current $J$, when
gcd$(k,n)=1$. We will explicitly give below the
corresponding $\psi$-matrix and {\it prove} it yields  a NIM-rep (as
always, the hardest thing to prove is that the entries $\N_{\la x}^y$ 
are non-negative integers). For such su($n$) level $k$, the
matrix $M[1]$ in \simcurM\ will be a modular invariant 
if and only if $n$ is odd. 
However, the matrices defined by the $\psi$-matrix given below define
a NIM-rep even if this is not the case. For example, if $n=2$ and $k$
is odd, they describe the `tadpole graph' of \refs{\bppz}. So for $n$
even, this NIM-rep generalises the tadpole.

Let $P_0$ be all level $k$ weights 
$\mu=(\mu_1,\ldots,\mu_{n-1})$ of su($n$), with ${n}$ dividing 
$t(\mu)=\sum_i i\mu_i$. Then $P_0$ will label both the rows and
columns of $\psi$. (We see explicitly in \simcurM\ that
$M[1]_{\mu\mu}\ne 0$ if and only if $\mu\in P_0$.) The matrix $\psi$
is actually the submatrix of the usual $S$-matrix, restricted to
$P_0$, and then rescaled by $\sqrt{n}$. More specifically, it is given
by 
\eqn\simple{
\psi_{\mu\nu} = c\, 
\exp\left[2\pi \i {t^+(\mu)\,t^+(\nu)\over (n+k)n}\right]  \,
\det \left[ \exp\left(-2\pi \i {\mu^+[i] \,\nu^+[j]\over n+k}\right)
\right]_{1\le i,j\le n}\,,}
where 
\eqn\cformula{
c=(n+k)^{{-n+1\over 2}} \i^{{n(n-1)\over 2}}\,,}
and as before $t^+(\mu)={n(n-1)\over 2}+\sum_ii\mu_i$, and 
$\mu^+[i]=n-i+\sum_{j=i}^{n-1}\mu_j$.

For any $\la\in P_+$ and $\mu,\nu\in P_0$, this NIM-rep coefficient 
$\N_{\la \mu}^\nu$ equals the fusion coefficient 
$N_{J^{i}\la\,\mu}^\nu$, where $i$ is the unique solution (mod $n$) to
the congruence $ki+t(\la)\equiv 0$. To see  this, note that this $i$
is the unique number for which $J^i\la\in P_0$. The $S$-matrix
factorises into  the tensor product of the matrix $\psi$ in \simple,
and a U(1) level $n$ $S$-matrix of phases. Now, whenever the
$S$-matrix decomposes into a tensor product, then the fusions
decompose into a product. So  the su$(n)$ level $k$ fusion 
$\la\times \mu$ here will be $J^{-i}((J^i\la) \times\mu)$, where the
fusion  $(J^i\la) \times\mu$ of weights in $P_0$ can be obtained  by
putting $\psi$ directly into Verlinde's formula \fusion.  This means
that \Nimdef\ here will equal $N_{J^{i}\la\,\mu}^\nu$, as desired. 

Hence we have {\it proven} that this choice of $\psi$ will
always define a  (non-negative integer) NIM-rep with the correct
exponents. A similar analysis applies for instance to so$(2n+1)$, 
so$(2n)$ and sp$(4n+2)$, when the level $k$ is odd.

\subsec{A case with fixed points}

Next, let us consider su($p$) with $p\geq 2$ prime, when $p$ divides 
the level $k$ (for $p=2$ we require $4$ to divide $k$). As we shall
see shortly, that the formula we are about to give is indeed
a NIM-rep follows from the conjectured formula in Section 6 of
\refs{\fsc} for the $S$-matrix $S^e$ of the chiral extension by a
simple current, as is also for example mentioned in \refs{\fhsw}. 
As long as $p^2$ does not divide $k+p$, we can
actually  {\it prove} that our $\psi$ matrix yields a (non-negative 
integer) NIM-rep. 

The exponents are the weights $\mu=(\mu_0;\mu_1,\ldots,\mu_{n-1})$
with $p$ dividing $t(\mu)=\sum_i i\mu_i$. There is a single fixed
point $\phi=\left({k\over p};{k\over p},\ldots,{k\over p}\right)$,  
which has multiplicity $p$ (as an exponent).  
The `boundary states' consist of all $J$-orbits $[\nu]$. Again, the
fixed point $\phi$ has multiplicity $p$. Although it may not be
completely obvious, these two sets always have the same cardinality,
namely  
\eqn\cardina{
{1\over p} \left({(n+p-1)!\over (n-1)!\, p!} +p^2 -1\right)\,.} 
Define $\psi_{\mu,[\nu]}$, when neither $\mu$ nor $\nu$ are 
fixed points by
\eqn\psione{
\psi_{\mu,[\nu]}=\sqrt{p}\, S_{\mu\nu}\,,}
where $S$ is the $S$-matrix of su($p$) level $k$. Of course this
formula is independent of which representative $\nu$ is taken for the
$J$-orbit $[\nu]$. 

\noindent When $\mu$ is the fixed point $\phi$ but $\nu$ is not, then 
\eqn\psitwo{\psi_{(\phi,i), [\nu]}=S_{\phi\nu}}
for any $i=0,\ldots,p-1$. 

\noindent When $\mu$ is not the fixed point, but $\nu=\phi$, then 
\eqn\psithree{\psi_{\mu,([\phi],j)}={S_{\mu\phi}\over \sqrt{p}}}
for any $j$.

\noindent Finally, when both $\mu=\nu=\phi$, we get for $p$ odd
\eqn\psifour{
\psi_{(\phi,h),([\phi],j)}= {1\over p} \,
\left(x-1+p\, \delta_{h,j}\right)\,,}
where 
\eqn\xf{
x=\left({p\over k+p}\right)^{{p-1\over 2}}\, s \qquad \hbox{for the
Legendre symbol}\qquad
s=\left({(k+p)/p \over p}\right)\,.}
Recall that the Legendre symbol $\left({(k+p)/p \over p}\right)$
equals  $0$ if $p^2$ divides $k+p$; otherwise it equals $\pm 1$ if
$(k+p)/p$ is or is not a perfect square mod $p$, respectively.
So $({a\over 3})=1,0,-1$ for 
$a\equiv 1,0,2$ (mod $3$), respectively, whereas  
$({a\over 5})=1,0,-1$ for $a\equiv \pm 1$, $a\equiv 0$,
$a\equiv \pm 2$ (mod 5), respectively. 

\noindent On the other hand, for $p=2$, the formula becomes
\eqn\psifive{
\psi_{(\phi,h),([\phi],j)}= {1\over 2}
\left[ (-1)^{{k\over 4}} \sqrt{{2\over k+2}}
+\i^{{k\over 4}} (-1)^{h+j} \right]\,. }
In this special case, the formula agrees with Eq.~(B.6) of
\refs{\bppz}. 

In order to calculate with and analyse this $\psi$ matrix, it is
convenient to use `fixed-point factorisation' \refs{\gw}. This  is a
way of computing $S$-matrix entries involving simple current fixed
points. For instance the entries $S_{\phi\nu}$ will equal either $0$
or  $\pm \left({p\over k+p}\right)^{{p-1\over 2}}$. The quantity $x$
in \xf\ equals $S_{\phi\phi}$. From this we quickly get that our
$\psi$ will indeed be unitary.

It is now straightforward to calculate the NIM-rep coefficients
$\N_{\la\, x}^y$ in terms of the WZW fusion rules.
We find (whenever $\mu,\nu$ are not fixed points)
\eqn\nimcoef{\eqalign{
{\cal N}_{\lambda\, [\mu]}^{[\nu]}=&\,\sum_J N_{J\lambda\, \mu}^\nu\cr
{\cal N}_{\lambda\, [\phi,i]}^{[\nu]}=&\,N_{\lambda\, \phi}^\nu\cr
{\cal N}_{\lambda\, [\phi,i]}^{[\phi,j]}=&\, {1\over p} \,
\left(N_{\lambda\, \phi}^\phi+(p\, \delta_{ij}-1)
{S_{\lambda\phi}\over S_{0\phi}} \right)\,.}}
Thus the first two are manifestly non-negative integers.
We know (from fixed-point factorisation) that the ratio
${S_{\lambda\phi}\over S_{0\phi}}$ here will be $0$ or $\pm 1$. So it 
suffices to verify, for all $\la\in P_+$, that 
\eqn\cond{
N_{\lambda\, \phi}^\phi\equiv {S_{\lambda\phi}\over S_{0\phi}}\ 
({\rm mod}\ p)\,.}

Condition \cond\  is certainly true when $p$ does not divide
$t(\lambda)$ --- in that case we have 
$N_{\lambda\, \phi}^\phi=0={S_{\lambda\phi}\over S_{0\phi}}$. When $p$ 
does divide $t(\lambda)$, then our 
${\cal N}_{\lambda [\phi,i]}^{[\phi,j]}$ 
precisely equals the extended fusion coefficient 
$N_{[\lambda] [\phi,i]}^{e\ [\phi,j]}$, for the conjectured 
$S$-matrix $S^e$ for the simple current chiral extension, as given in
\fsc. This observation is also made in \refs{\fhsw}. 
So provided their conjecture \refs{\fsc} is correct (as is expected),
then the $\psi$ given above does indeed define a (non-negative
integer) NIM-rep. Some relation between NIM-reps and extended
$S$-matrices is expected --- see {\it e.g.}\ section 6 of
\refs{\fsgen}, Thm.~4.16 of \refs{\bek}, and (B.6) in \refs{\bppz}.  

Assume now that $p^2$ does not divide $p+k$. Then $p$ does not divide
the integer ${1\over S_{0\phi }^2}=\left({k\over p}+1\right)^{p-1}$. 
So it suffices to show that $p$ must divide the integer
\eqn\conn{{1\over S_{0\phi}^2}\,
\left(N_{\lambda\, \phi}^\phi-{S_{\lambda\phi}\over S_{0\phi}}\right)=
\sum_{\mu} 
{S_{\lambda\mu}\, S_{\phi\mu}^2 \over S_{0\mu}\,S_{0\phi}^2} +  
{S_{\lambda\phi}\over S_{0\phi}}
\left({S_{\phi\phi}^2\over S_{0 \phi}^2}
-{1\over S_{0\phi}^2}\right)\,,} 
where the sum is over all non-fixed-points $\mu$.

\noindent Consider the sum
\eqn\conone{\sum_\mu 
{S_{\lambda \mu}\, S_{\phi\mu}^2\over S_{0\mu}\, S_{0\phi}^2}=
p\sum_{[\mu]}{S_{\lambda\mu}\over S_{0\mu}}\,,}
where the first sum is over all non-fixed-points $\mu$, and the second
sum is over all $J$-orbits of non-fixed-points $\mu$ with
$S_{\mu \phi}\ne 0$. Note that this sum over $[\mu]$ will be an
algebraic integer (since each ratio ${S_{\lambda\mu}\over S_{0\mu}}$ 
is an algebraic integer, being the eigenvalue of an integer matrix)
and it will be rational (since it will be fixed by all
Galois automorphisms), and therefore it must be an
integer. So $p$ divides the first term on the right hand side of
\conn. 

\noindent Also, Fermat's Little Theorem implies that $p$ must divide
the integer  
\eqn\contwo{{S_{\phi\phi}^2\over S_{0\phi}^2}-
{1\over S_{0\phi}^2}=1-\left({k\over p}+1\right)^{p-1}\,.}
Thus $p$ divides \conn, and hence ${\cal N}$ will indeed be a
non-negative integer, provided only that $p$ does not divide 
${k\over p}+1$.

\newsec{Conclusion}

\noindent In this paper we have made a proposal for the boundary
states of the WZW models that correspond to the conjugation modular
invariant associated to a symmetry $\omega$ of the unextended Dynkin
diagram. The boundary states of this theory are labelled by 
the $\omega$-twisted representations, and they are described by a
formula \bounfinal\ similar to Cardy's expression for the boundary
states of the diagonal theory. The corresponding NIM-rep agrees
precisely with the twisted fusion rules. These fusion rules are   
given by a Verlinde-like formula \argumenttwo.

We have given explicit formulae for the $\psi$-matrix for all possible 
cases (section~4), and we have related the NIM-rep entries 
to untwisted fusion rules, thereby proving their
integrality. We have
also checked nonnegativity explicitly for numerous cases. Some of
the corresponding fusion graphs have been collected in the appendix. 
\smallskip

Our formulae seem to generalise further to certain other classes of
modular invariants, in particular those associated to simple
currents. In order to illustrate this point, we have described two
classes in detail in section~6. The results presented there already
suffice to determine the NIM-reps for all simple current modular
invariants for su$(p)$ level $k$, when $p\geq 3$ is prime. We have
also proven that our NIM-rep coefficients are non-negative integers
(at least when $p^2$ does not divide $k+p$). The general
case (that will be described elsewhere) will essentially cover all
remaining modular invariant WZW models. 

\vskip1cm

\centerline{{\bf Acknowledgements}}\par\noindent

\noindent We are grateful to David Evans, Peter Goddard, Valya
Petkova, Andreas
Recknagel and Jean-Bernard Zuber for useful conversations and
communications. 
We thank J\"urgen Fuchs, Christoph Schweigert, and Bert Schellekens
for explaining their work to us. We also thank Thomas Mettler for
drawing our attention to a number of small mistakes in the published
version which are corrected here.   
MRG is grateful to the Royal Society for a 
University Research Fellowship.  He also acknowledges partial support
from the EU network `Superstrings' (HPRN-CT-2000-00122), as well as
from the PPARC special grant `String Theory and Realistic Field
Theory', PPA/G/S/1998/0061. TG warmly thanks St. John's College,
Cambridge, which generously supported him in the early stages of  this
project; his  research is supported in part by NSERC.

\vskip1.5cm

\appendix{A}{Explicit descriptions of the fusion graphs}

In this appendix we give explicit descriptions of some of the NIM-reps
we have found in terms of the corresponding graphs. For the examples
under consideration, the NIM-reps are often characterised in 
terms of the matrix associated to the field $\lambda=\Lambda_1$. The 
corresponding graph has $M$ vertices labelled by the rows (or columns)
of $\N_\lambda$, and the vertex associated to $i$ and $j$ are linked
by $(\N_{\lambda})_{ij}$ lines. Incidentally, given our discussion of
section~3 above, these graphs are precisely the fusion graphs that
describe the fusion of the field $\lambda$ with the twisted
representations (that label the rows and columns of $\N_\lambda$). In
particular, the NIM-rep matrices considered here are therefore
symmetric, and the relevant graphs are not oriented.

For the case of su$(n)$ with $\omega$ being charge conjugation, the
graphs are explicitly given in Figure~1. The left-right symmetry of
the su(4) graphs is due to the symmetry $J^B$ of the
$A_3^{(2)}=D_3^{(2)}$ Dynkin diagram. In addition, the graphs for
su(4) are 2-colourable because of the symmetry $J^E$ of the
$D_3^{(2)}$ diagram. Finally, there is also a left-right symmetry of
the su(3) and su(5) graphs when the level is odd, due to the
$C_1^{(1)}=A_1^{(1)}$ and $C_2^{(1)}$ Dynkin symmetries $J^B$,
respectively.  

\ifig\sunpic{The NIM-rep graphs for su$(N)$ with charge conjugation
for $N=3,4,5$ and $k=1,2,3$.}
{\epsfxsize4.5in\hskip.2cm\epsfbox{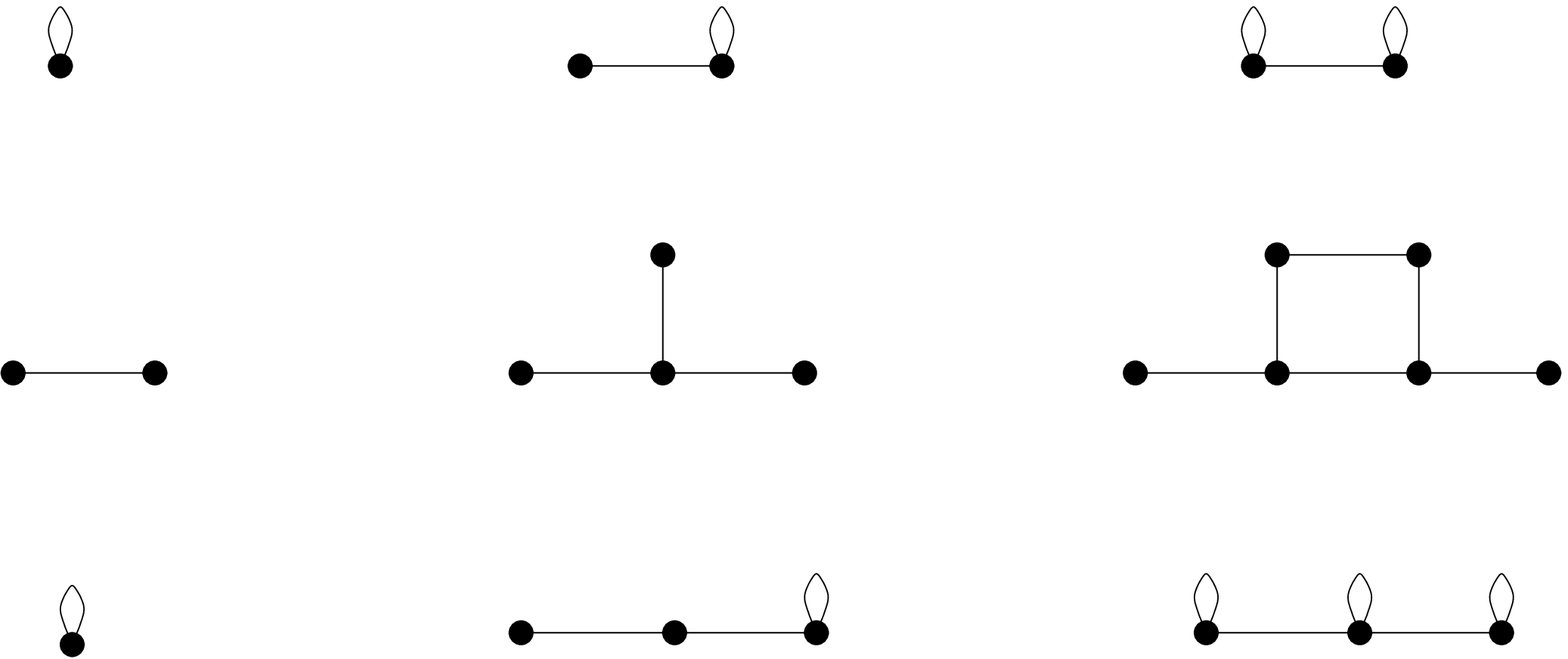}}

\noindent For the case of so$(2N)$ with chirality flip, the
relevant graphs are explicitly given in Figure~2. The graphs for
so$(6)$ do not agree with those of su$(4)$ given in Figure~1 since,
under the identification su$(4)\cong$ so$(6)$, $\Lambda_1$ of
su$(4)$ corresponds to $\Lambda_3$ of so$(6)$. All of these graphs
have an order-$2$ symmetry, which is due to the symmetry $J^B$ of the
$D_3^{(2)}$ and $D_4^{(2)}$ Dynkin diagrams. The graphs are
disconnected because of the symmetry $J^E$ of the $A_3^{(2)}$ and
$A_4^{(2)}$ diagrams. In particular $Q^E(\L_1)=0$, and thus the
grading \scnim\ implies that each component consists of the $a\in\B$
with a fixed value of the grade $Q^E(a)$.  

\ifig\sonpic{The NIM-rep graphs for so$(2N)$ with chirality flip
for $N=3,4$, $k=1,2,3$.} 
{\epsfxsize4.5in\hskip0.1cm\epsfbox{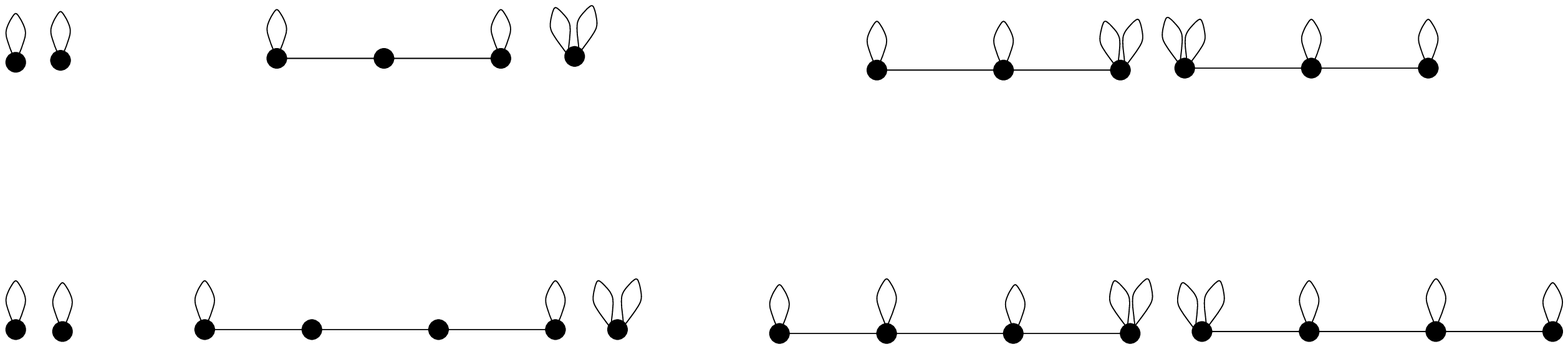}}

\noindent For the case of so$(8)$ with triality, and the case of $E_6$
with charge conjugation, the relevant graphs are explicitly given in
Figure~3 and Figure~4. respectively.  
\ifig\trialitypic{The NIM-rep graphs for so$(8)$ with triality for
$k=2,3,4,5,6,7$.} 
{\epsfxsize4.5in\hskip-.5cm\epsfbox{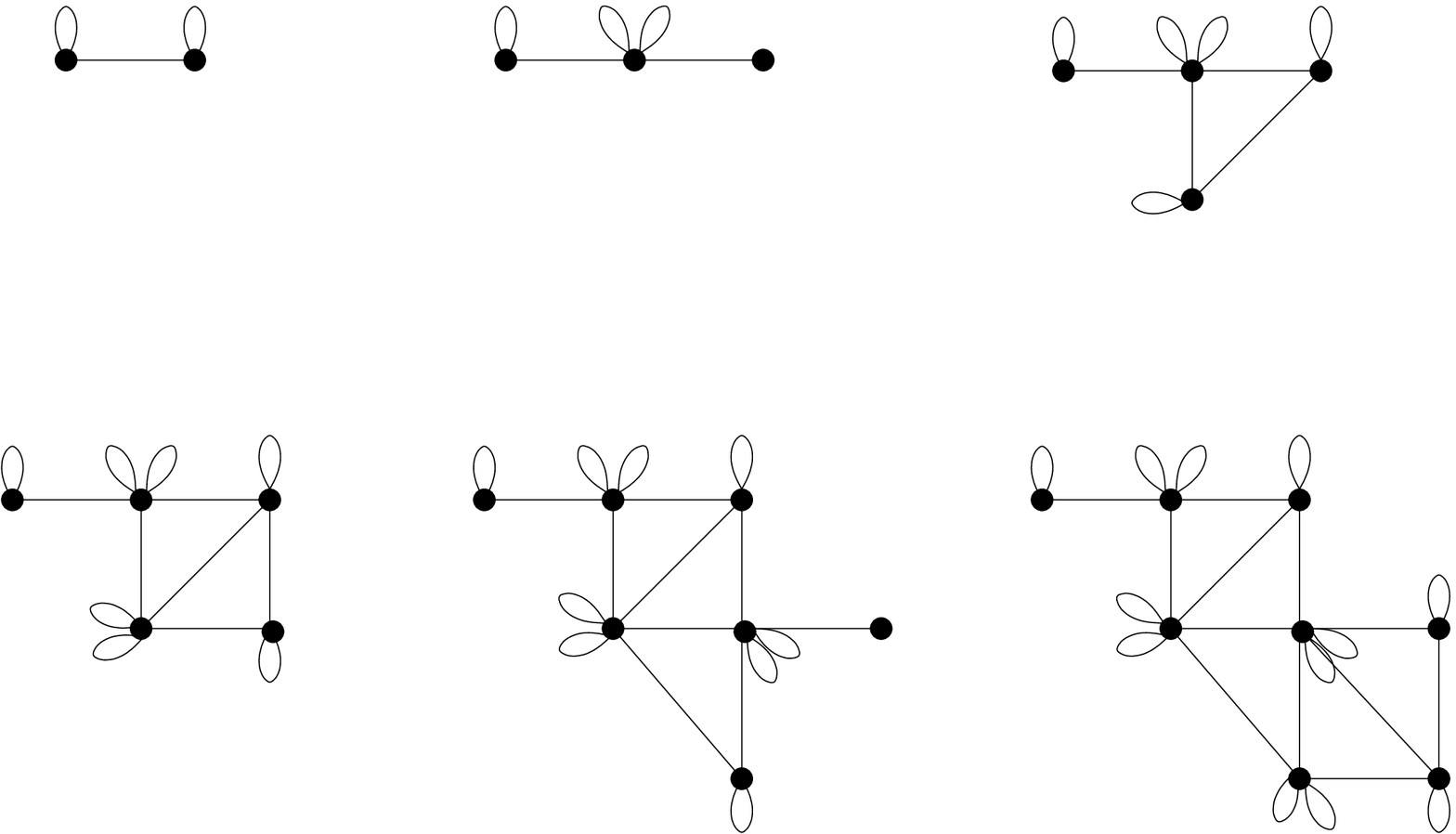}}

\ifig\esixpic{The NIM-rep graphs for $E_6$ with charge conjugation for 
$k=2,3,4$.} 
{\epsfxsize4.5in\hskip-.5cm\epsfbox{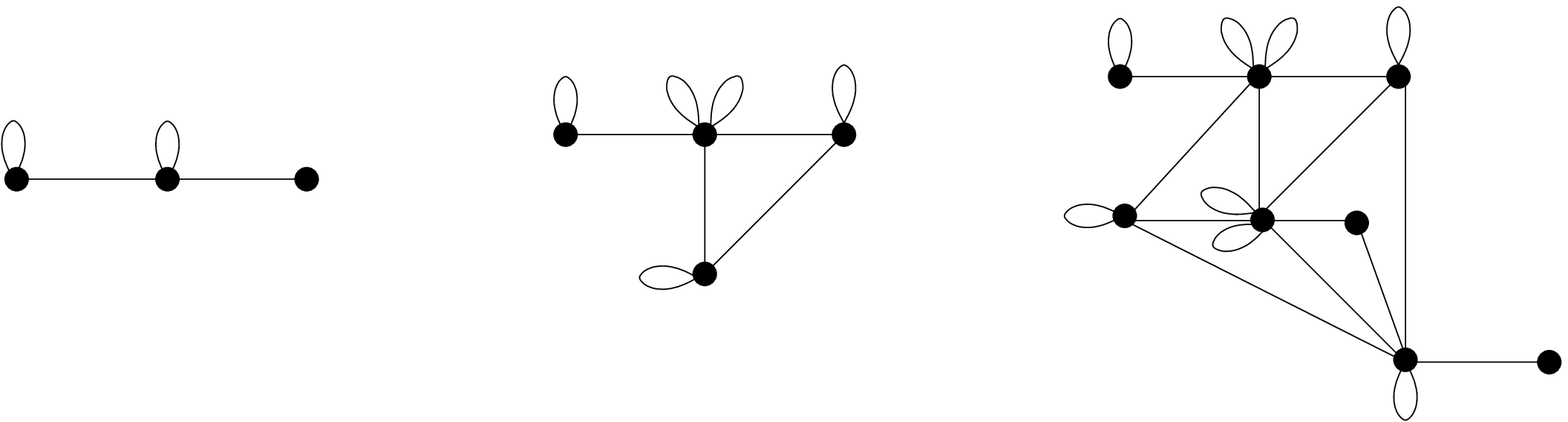}}

\listrefs

\end